\documentclass[fleqn,usenatbib]{mnras}

\usepackage{newtxtext,newtxmath}

\usepackage[T1]{fontenc}
\usepackage{ae,aecompl}
\usepackage{xcolor}
\usepackage{ulem}
\usepackage{soul}
\usepackage{ragged2e}

\usepackage{subfig}

\usepackage{graphicx}	
\usepackage{amsmath}	
\usepackage{verbatim}
\usepackage{bm}
\usepackage{multirow}

\usepackage{booktabs}

\usepackage{tablefootnote}

\graphicspath{{./plots/}}

\newcommand{\Mwd}{$M_{\mathrm{WD}}$}

\newcommand{\Msun}{$M_{\odot}$}

\defcitealias{cunningham19}{CU19}

\makeatletter
\DeclareRobustCommand{\I}{%
	\mbox{\check@mathfonts\fontsize\sf@size\z@\selectfont I}%
}
\DeclareRobustCommand{\V}{%
	\mbox{\check@mathfonts\fontsize\sf@size\z@\selectfont V}%
}
\makeatother

\title[Two new magnetic period bouncers]{Discovery of two new polars evolved past the period bounce}

\author[T. Cunningham et. al]
{Tim Cunningham$^{1}$\thanks{E-mail: tim.cunningham@cfa.harvard.edu }\thanks{NASA Hubble Fellow.},
Ilaria Caiazzo$^{2,3}$ ,
Gracjan Sienkiewicz$^{4}$,
Peter J. Wheatley$^{4,5}$,
{Boris T. G\"ansicke$^{4,5}$,} 
\newauthor{Kareem El-Badry$^{3}$,}
Riccardo Arcodia$^{6}$,
Dave Charbonneau$^{1}$,
Liam Connor$^{1}$,
Kishalay De$^{6}$,
{Pasi Hakala$^{7}$,}
\newauthor{Scott J. Kenyon$^{1}$,}
{Sumit Kumar Maheshwari$^{8}$,}
Antonio C. Rodriguez$^{3}$,
Jan van Roestel$^{9}$,
\newauthor{Pier-Emmanuel Tremblay$^{4,5}$}
\\
$^{1}$Center for Astrophysics | Harvard \& Smithsonian, 60 Garden St., Cambridge, MA 02138, USA\\
$^{2}$Institute of Science and Technology Austria, Am Campus 1, 3400 Klosterneuburg, Austria\\
$^{3}$Division of Physics, Mathematics and Astronomy, California Institute of Technology, Pasadena, CA91125, USA\\
$^{4}$Department of Physics, University of Warwick, Coventry, CV4 7AL, UK\\
$^{5}$Centre for Exoplanets and Habitability, University of Warwick, Gibbet Hill Road, Coventry CV4 7AL, UK\\
$^{6}$MIT-Kavli Institute for Astrophysics and Space Research, 77 Massachusetts Avenue, Cambridge, MA 02139, USA\\
$^{7}$Finnish Centre for Astronomy with ESO (FINCA), Quantum, University of Turku, FI-20014, Finland\\
$^{8}$Hamburger Sternwarte, University of Hamburg, Gojenbergsweg 112, 21029 Hamburg, Germany\\
$^{9}$Anton Pannekoek Institute for Astronomy, University of Amsterdam, 1090 GE Amsterdam, The Netherlands
}

\date{Accepted XXX. Received YYY; in original form ZZZ}

\pubyear{2023}

\begin{document}
\label{firstpage}
\pagerange{\pageref{firstpage}--\pageref{lastpage}}
\maketitle

\begin{abstract}
We report the discovery of two new magnetic cataclysmic variables with brown dwarf companions and long orbital periods ($P_{\rm orb}$\,=\,$95\pm1$ and $104\pm2$ min). This discovery increases the sample of candidate magnetic period bouncers with confirmed sub-stellar donors from four to six. 
We also find their X-ray luminosity 
from archival \textit{XMM-Newton} observations to be in the range $L_{\rm X}$\,$\approx$\,$10^{28}$--$10^{29}$\,$\mathrm{erg\,s^{-1}}$ in the 0.25--10\,keV band. This low luminosity is comparable with the other candidates, and at least an order of magnitude lower than the X-ray luminosities typically measured in cataclysmic variables. The X-ray fluxes imply mass transfer rates that are much lower than predicted by evolutionary models, even if some of the discrepancy is due to the accretion energy being emitted in other bands, such as via cyclotron emission at infrared wavelengths. Although it is possible that some or all of these systems formed directly as binaries containing a brown dwarf, it is likely that the donor used to be a low-mass star and that the systems followed the evolutionary track for cataclysmic variables, evolving past the  period bounce. The donor in long period systems is expected to be a low-mass, cold brown dwarf. This hypothesis is supported by near-infrared photometric observations that constrain the donors in the two systems to be brown dwarfs cooler than $\approx$1100\,K (spectral types T5 or later), most likely losing mass via Roche Lobe overflow  or winds. The serendipitous discovery of two magnetic period bouncers in the small footprint of the XMM-newton catalog implies a large space density of these type of systems, possibly compatible with the prediction of 40--70 per cent of magnetic cataclysmic variables to be period bouncers.

\end{abstract}




\section{Introduction}
\label{sec:intro}

Cataclysmic variables (CVs) are binary stars in which a white dwarf accretes from a low-mass companion, typically a late-type main-sequence star or a sub-stellar companion. 
In the standard model of CV evolution, CVs with a main-sequence companion evolve toward shorter periods because of angular momentum losses due to a combined effect of magnetic braking and gravitational wave emission, following a track that is mostly determined by the mass of the stellar donor \citep[e.g.][]{knigge2011}. At longer periods, magnetic braking is thought to dominate angular momentum losses, while at periods below 2 hours, the dominant contribution is thought to be from gravitational wave emission, although there is evidence that residual magnetic braking plays an important role \citep{belloni2020}. 
As the CV reaches the period minimum \citep[at about 80 minutes,][]{faulkner1971,paczynski1981,knigge2011} and the donor continues to lose mass, the thermal timescale of the donor becomes larger than the timescale of mass loss. This means that the donor, no longer in thermal equilibrium, ceases to shrink as it loses mass. As the donor becomes increasingly degenerate, its radius increases in response to mass loss, and the system evolves back towards longer orbital periods to accommodate the larger donor. It 
is predicted that there should be a large number of ``period-bounce'' CVs in the Galaxy, with estimates varying from 40--70\% of the whole CV population \citep{kolb1993,goliasch2015,belloni2020,pala2022}.  
There have been very few period-bouncers discovered so far, however, probably due to the fact that they show lower accretion rates (and by proxy a lower X-ray luminosity, fainter accretion disks, and rarer outbursts) and their companions are faint sub-stellar objects, which are more difficult to detect.
There are now close to two thousand CVs known of various types \citep{ritter2003,inight2023}, but only about 25 have been confirmed to be period bouncers \citep{munoz-giraldo2024-period-bounce,guidry2021}. From a volume-limited sample of CVs within 150\,pc, \citet{pala2020} estimated the period-bounce population accounts for some 7--14 per cent of the total observed CV population. Thanks to eROSITA, this volume-limited sample has been recently updated, including a misclassified object and a new discovery, bringing the fraction close to 25 per cent, reducing the discrepancy with theoretical predictions \citep{rodriguez2024}. It has been recently suggested that the discrepancy between the observed and predicted fraction of bouncers might be due to the fact that in many CVs the white dwarf could become magnetic after the bounce. In this scenario, after the magnetic field emerges, the connection between the field of the white dwarf and that of the secondary would cause the system to detach and accretion to stop. The predicted number of period bouncers would therefore be reduced by 60--80 per cent \citep{schreiber2023}. This prediction hinges on the assumption that strong magnetic fields in white dwarfs can either be created \citep[as for example through the crystallization dynamo,][]{isern2017} or emerge to the surface \citep[as for example through diffusion of fossil fields,][]{camisassa2024} at white dwarf cooling ages of a few Gyr. 

White dwarfs have been detected to have magnetic fields ranging from a few kG to hundreds of MG \citep{ferrario2015,bagnulo2019} and at least 20\% of isolated white dwarfs have detectable magnetic fields \citep{bagnulo2020}. Magnetic fields in accreting systems, if strong enough, can disrupt the flow in the accretion disk, funneling infalling material along field lines to the magnetic pole, or poles, of the white dwarf. CVs containing magnetic white dwarfs are usually divided in two categories depending on the strength of the field and on the relation between the spin period of the white dwarf and the orbital period. When the field is lower than a few MG and the spin period of the white dwarf is not synchronized with the orbital period, the system is called an intermediate polar (or IP). In these type of systems, accretion tends to spin up the white dwarf; also, they usually show the presence of a truncated disk. In CVs with larger fields ($B$\,$\gtrsim$\,10\,MG), called polars, the disk is often completely disrupted and the material flows directly from the first Lagrange point ($L_1$) along magnetic field lines to the surface of the white dwarf. In these systems, torques between the magnetic moment of the white dwarf and that of the companion star tend to synchronize the rotation period of the white dwarf with that of the orbit. Magnetic CVs with field strengths $B\gtrsim1$\,MG have been found in population studies to account for approximately one third of the total CV population \citep{pala2017,inight2023}. Among the 25 confirmed or candidate period-bounce CVs \citep{munoz-giraldo2024-period-bounce,guidry2021,kawka2021,rodriguez2024}, only 7 have a detected magnetic field (see Table\,\ref{tab:pbs}). It has been predicted from simulations that the number of magnetic period-bouncers may be lower by 10\% compared to their non-magnetic counterparts \citep{belloni2020}, but this is still at odds with the current population, unless selection effects disfavor magnetic period bouncers.

Low accretion rates ($\lesssim 10^{-13}$~M$_\odot$) are observed in three different types of magnetic CVs: polars in a transient low-accretion state, pre-polars and period bouncers. The former are normal polars, i.e. polars with a stellar companion filling its Roche lobe, that are sometimes observed to transition to prolonged states (days to years) with significantly lower accretion rates than expected for Roche lobe overflow \citep{kafka09,schwope02}. The reason behind this transition is poorly understood; it has been suggested that magnetic spots on the donor are moving past $L_1$, interrupting the flow \citep{livio94}, but this explanation is hard to reconcile with the extended low states observed in some polars. An example is the polar EF Eri, which was in a low state for almost 26 years between 1997 and late 2022 \citep{wheatley98,howell06,filor2024} (EF Eri could also be a period bouncer, see section \ref{sec:pop}). 
Many systems that were initially interpreted as low-state polars \citep[e.g.][]{reimers1999,reimers2000,schmidt2005} were later shown to be pre-polars \citep{tout2008,schwope2009}: magnetic white dwarfs whose stellar donor is underfilling its Roche lobe. In these systems, the low accretion rate is due to the fact that accretion is driven by winds from the stellar companion, rather than Roche lobe overflow, and the companion main sequence star is usually clearly detected. This mechanism produces a lower mass loss rate from the donor. These systems are often detected thanks to their strong cyclotron emission, often at optical wavelengths \citep[see ][ and references therein]{parsons2021,vanroestel2024}. Pre-polars might be magnetic CVs in which Roche-lobe-filling accretion has not started yet; alternatively, it has been suggested that pre-polars are CVs in which the white dwarf has recently become magnetic due to the so-called crystallization dynamo \citep{isern2017}, and the connection between the white dwarf's magnetic field with that of the secondary star has caused the binary to detach, due to synchronization torques and reduced angular momentum loss \citep{schreiber2021}. 

Finally, low accretion rates are expected in polars that have evolved past the period bounce: in these systems, angular momentum losses, which are thought to be driven only by weak gravitational wave emission, lead to a slow increase in the orbital period and a low mass-transfer rate, similar to the non-magnetic period-bouncers \citep[e.g.][]{pala2018,inight2023}. Other than EF Eri, a few candidate magnetic period bouncers are known with periods close to the period minimum, low accretion rates and possible sub-stellar companions \citep{breedt2012}: SDSSJ151415.65+074446.5, SDSS J125044.42+154957.4, and V379 Vir (see also section\,\ref{sec:pop}). These three candidate period-bouncers were confirmed to be currently accreting owing to the detection of X-rays in XMM-Newton and eROSITA observations \citep{munoz-giraldo2023}. Also discovered via X-ray emission in eROSITA, 1eRASS J054726.9+132649 was confirmed by \citet{rodriguez2024} as a magnetic CV evolved past the period minimum to a period of 94\,min. SMSS\,J1606-1000 was recently discovered to have a sub-stellar companion and indication of ongoing accretion \citep{kawka2021}, but no X-ray observations are available for the system. Finally, \citet{guidry2021} discovered a candidate magnetic period bouncer with a significantly longer orbital period of 2 hours and strong cyclotron features detected at optical wavelengths. This system also has not been targeted by X-ray observations, but the presence of cyclotron emission strongly suggests that accretion is taking place. Collectively, these systems yield a sample of seven magnetic, candidate period-bounce CVs, four of which have a confirmed sub-stellar companion (we present the previously known candidates in section \ref{sec:pop}).

We here report the serendipitous discovery of two new magnetic period bouncers with very low accretion rates and long orbital periods (95--104\,mins) identified in the XMM-Newton source catalog \citep[the 4XMM-DR13 release;][]{webb2020-4XMM}. In Section\,\ref{sec:discovery}, we describe the XMM observations as well as the follow-up optical spectroscopy and photometry. In Section\,\ref{sec:analysis}, we describe the observational constraints on the systems' parameters, including the physical properties of the white dwarfs, the nature of the companion stars, the orbital geometries and accretion rates. In Section\,\ref{sec:discussion}, we discuss our interpretation of the nature of the systems, possible evolution scenarios and the implications of the discovery on the population of period bouncers. In Section\,\ref{sec:conclusions}, we summarise our results.

\begin{figure}
	\centering\includegraphics[width=0.8\columnwidth]{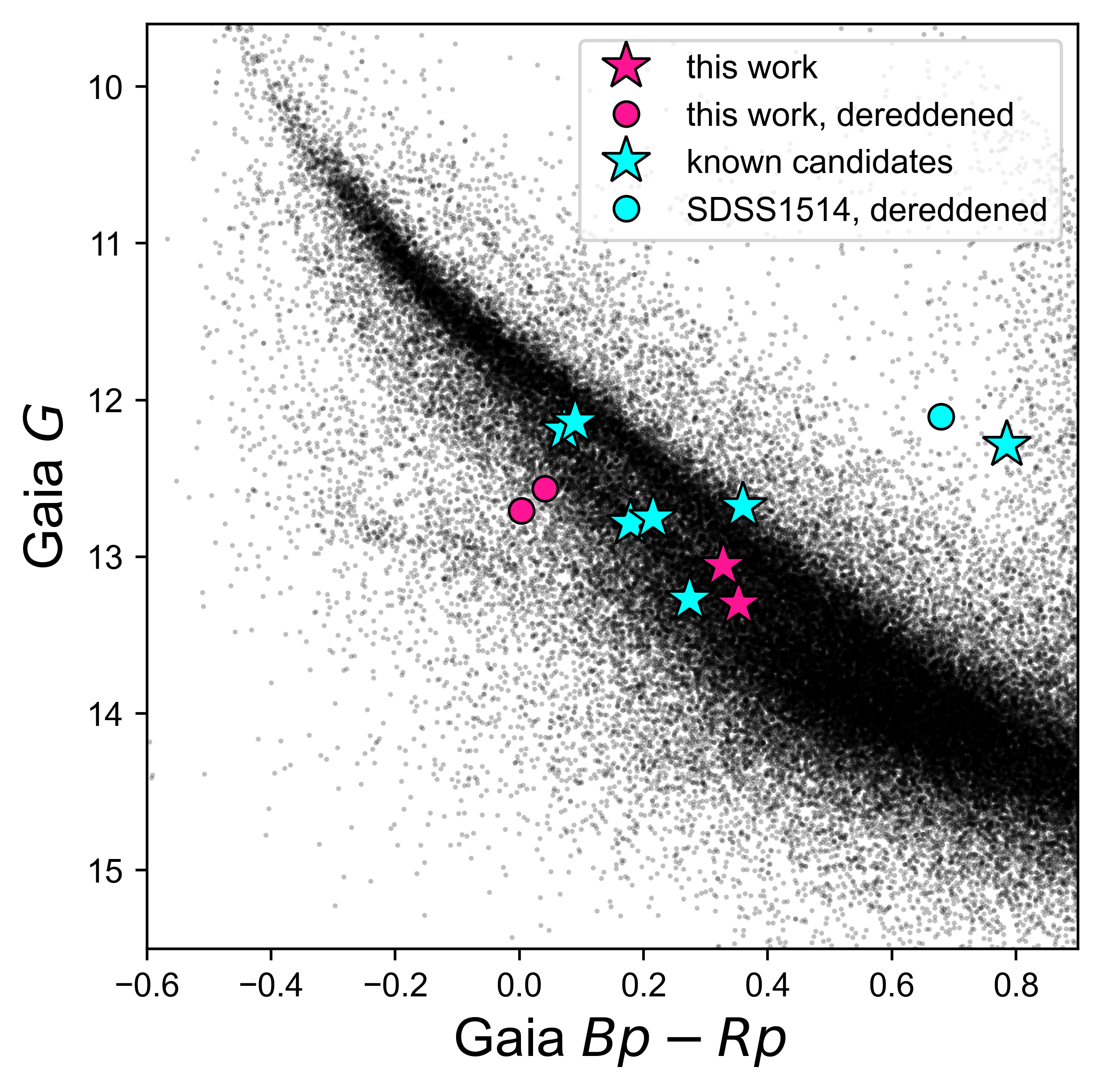}
	\caption{\textit{Gaia} color-magnitude position of the two systems presented in this study (the magenta star markers) and of the previously known candidate magnetic period bouncers (in cyan, see Table\,\ref{tab:pbs}). Extinction at the location of the systems was obtained from the Bayestar19 dust map \citep{green2019}, and the de-reddened location is shown for the two systems in this work and for SDSS\,1514, the only ones to have appreciable extinction, as round markers. The de-reddened Gaia colours and magnitudes of the two systems in this work are very similar and consistent with white dwarfs at around 10\,000\,K. The location of SDSS\,1514 far to the right of and above the white dwarf cooling sequence is due to the strong cyclotron emission in the $R_p$ and $G$ filters.}
	\label{fg:GaiaCMD}
\end{figure}

\section{Discovery and Observations}
\label{sec:discovery}
\subsection{Discovery in the 4XMM-DR13 Source Catalogue}

Here we report the serendipitous discovery of two new magnetic CVs: WD\,J182047.71-042253.08 (Gaia DR3 4269487247901922176, hereafter WD\,J1820) and WD\,J190700.40+205226.96 (Gaia DR3 4519789940386173696, hereafter WD\,J1907). The two systems were identified as part of a search for soft X-ray emission from likely isolated white dwarfs in the XMM-Newton source catalog \citep{webb2020-4XMM} that will be published in a companion article. We performed a cross-match between high-probability white dwarf candidates \citep{ngf19,ngf21} identified in \textit{Gaia} DR3 \citep{gaia2018} and the XMM-Newton source catalog, and followed up with optical spectroscopy the best candidate X-ray emitting white dwarfs that lacked a previous spectroscopic identification. 

\begin{figure}
	\centering
    \subfloat{\includegraphics[width=\columnwidth]{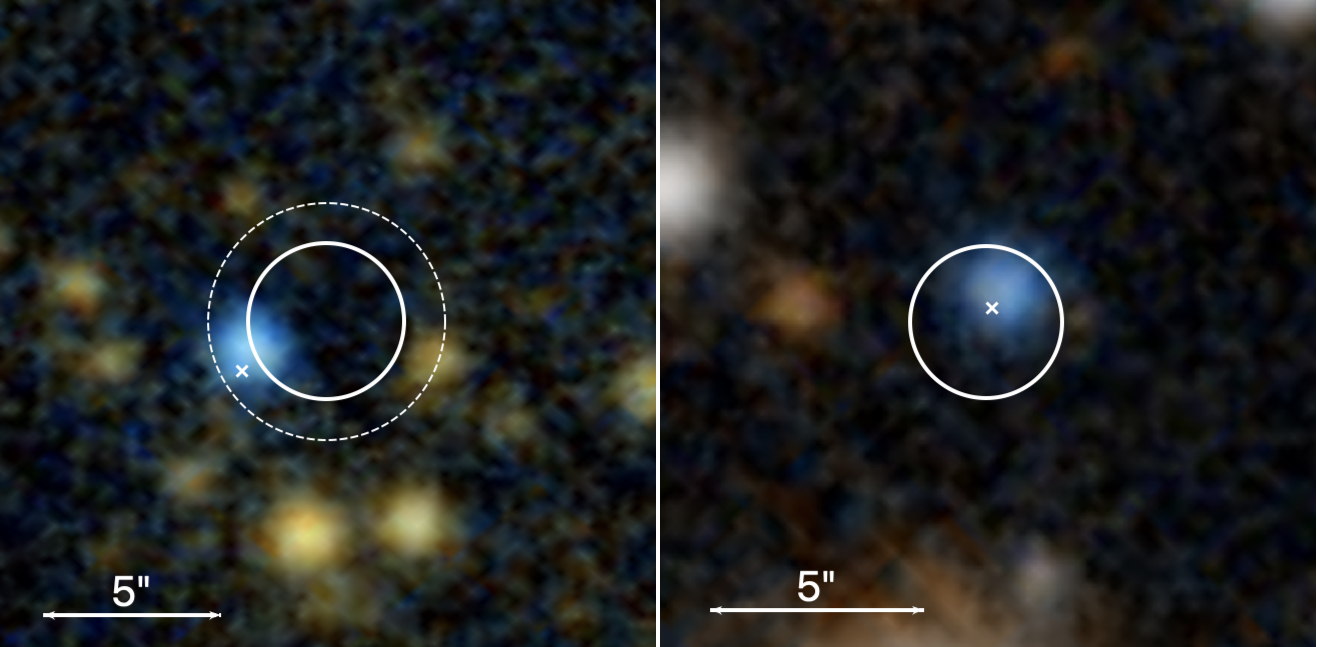}}
	\caption{PanSTARRS DR1 colour image with bands $g$ and $z$ for WD\,J1820 (left) and WD\,J1907 (right). From \textit{Gaia} astrometry, the sky position of the white dwarf at the mean epoch of the XMM-Newton observations is indicated with the white cross. The solid white circle is centered on the 4XMM-DR13 source position, with a radius equal to the 68\% confidence radius on the source position, as provided in the 4XMM-DR13 catalogue. In the left panel, the dashed white circle indicates the 90\% confidence radius.}
	\label{fg:PS1-4XMM-pos}
\end{figure}

WD\,J1820 and WD\,J1907 appear as blue and faint objects and therefore as fairly massive, isolated white dwarfs in the Gaia color-magnitude diagram (see Fig.~\ref{fg:GaiaCMD}) and are coincident (within 4'') with two soft X-ray sources in the XMM catalog (see Fig.\,\ref{fg:PS1-4XMM-pos}). The source coincident with WD\,J1820 was detected serendipitously via one XMM-Newton observation (obsid: 0821890101; PI: Mereghetti) on the 24$^{\rm th}$ of September 2018 with an exposure time of 33\,ks, targeting the pulsar PSR\,B1818-04. The other system studied in this work, WD\,J1907, was detected in two XMM-Newton observations on the 27$^{\rm th}$ of September 2019 (obsid: 0840843401; PI: Stelzer) and on the 20$^{\rm th}$ of September 2020 (obsid: 0860303001; PI: Stelzer) with exposure times of 32.4 and 31.7\,ks, respectively, targeting the nearby early M-dwarf PM\,I19072+2052. For WD\,J1820, the source catalogue reports a total of 51 source counts across all three EPIC cameras, whilst for the two observations of WD\,J1907 in 2019 and 2020, the source counts were 75 and 59, respectively. The source counts in each camera and total count rates are provided in Table\,\ref{tab:4XMM}.
For WD\,J1907, simultaneous observations to the EPIC X-ray cameras were obtained in the UV with the Optical Monitor (see sec.\,\ref{sec:photometry}). After correcting for proper motion, the sky separation between the Gaia position and X-ray source position is $3\pm 2$'' and $0.4\pm1.7$'' for WD\,J1820 and WD\,J1907, respectively. 

Fig.\,\ref{fg:PS1-4XMM-pos} shows a PanSTARRS DR1 color image of WD\,J1820 and WD\,J1907 on the left and right, respectively. The 4XMM-DR13 source position is indicated by the white circle, which is centered on the catalog sky position and has a radius equivalent to the 68\% confidence radius on the source position as defined by the source catalog. The sky position of the two Gaia white dwarf candidates at the time of the XMM-Newton observations is shown as white crosses (the position has been corrected for proper motion to the mean epoch of the XMM-Newton observations). We find that WD\,J1907 is consistent to within 1$\sigma$, whilst WD\,J1820 is well within the 90\% confidence interval on the source position. Given that the two systems presented here both exhibit multiple additional observational signatures associated with accretion -- namely cyclotron emission, narrow emission lines, and binary companions -- the X-ray association is sufficiently robust to confirm ongoing accretion in the systems.

\subsection{XMM-Newton Observations}

\begin{table}
	\centering
	\caption{X-ray source counts for WD\,J1820 and both observations of WD\,J1907 as reported in the 4XMM-DR13 serendipitous source catalogue for the 0.2--12\,keV band. We include the source counts report for the three EPIC cameras (PN, MOS1 and MOS2), as well as the combined EPIC counts and count rates.}
	\label{tab:4XMM}
	\begin{tabular}{lrrr}
		\hline        
		 &  WD\,J1820 & WD\,J1907 & WD\,J1907\\         		\hline
\vspace{3pt}

obsid           &    0821890101  &  0840843401 & 0860303001  \\ 

PN counts	    & $27\pm9$		     & $51\pm10$ & $36\pm9$ \\
M1 counts	& $9\pm5$		     & $17\pm5$ &  $12\pm6$ \\ \vspace{3pt}
M2 counts   & $15\pm7$		     & $8\pm5$&  $11\pm5$ \\
EPIC counts	    & $51\pm12$		     & $75\pm13$&  $59\pm12$ \\
EPIC count rate  & \multirow{2}{*}{$2.8 \pm 0.7$} & \multirow{2}{*}{$7.5\pm1.3$} &  \multirow{2}{*}{$4.4 \pm 0.9$} \\

[$10^{-3}$\,cts/s] & & & \\

		\hline
	\end{tabular}
\end{table}
From the XMM-Newton Science Archive \citep{xmm-xsa}, we acquired the observation data files (ODF) for the EPIC cameras PN and MOS (M1 and M2). The sky position of the two sources fell onto the chip for all three EPIC cameras in all three observations (one observation for WD\,J1820, and two for WD\,J1907). From our analysis, we exclude the first PN exposure of WD\,J1907 after visual inspection reveals the source fell near the edge of one of the PN chips, and suffers from contamination. The EPIC data was reduced using v21.0.0 of the Science Analysis Software (SAS) designed for reduction of XMM-Newton data \citep{gabriel2004-sas}. We filtered the event files using the routine \texttt{evselect} and Good Time Intervals (GTI) that were defined with standard defaults and thresholds on the count rate of 0.6\,counts/s for PN and 0.35\,counts/s for MOS, chosen by visual inspection of the each observation. This procedure resulted in total effective exposure times of 21.0\,ks (PN) and 28.2\,ks (M1+M2) for WD\,J1820, and 27.1\,ks (PN) and 40.2\,ks (M1+M2) for WD\,J1907. We extracted spectra for the sources and adjacent background regions using \texttt{evselect}, with background regions placed on the same chip as the sources. The source and background aperture radii were 20\,arcsec and 90\,arcsec, respectively. The X-ray spectral analysis is presented in Section\,\ref{sec:results-xray-fits}.

\subsection{Keck/LRIS Spectroscopy}
\label{sec:photometry}
\begin{figure*}
	\centering
	\includegraphics[width=\textwidth]{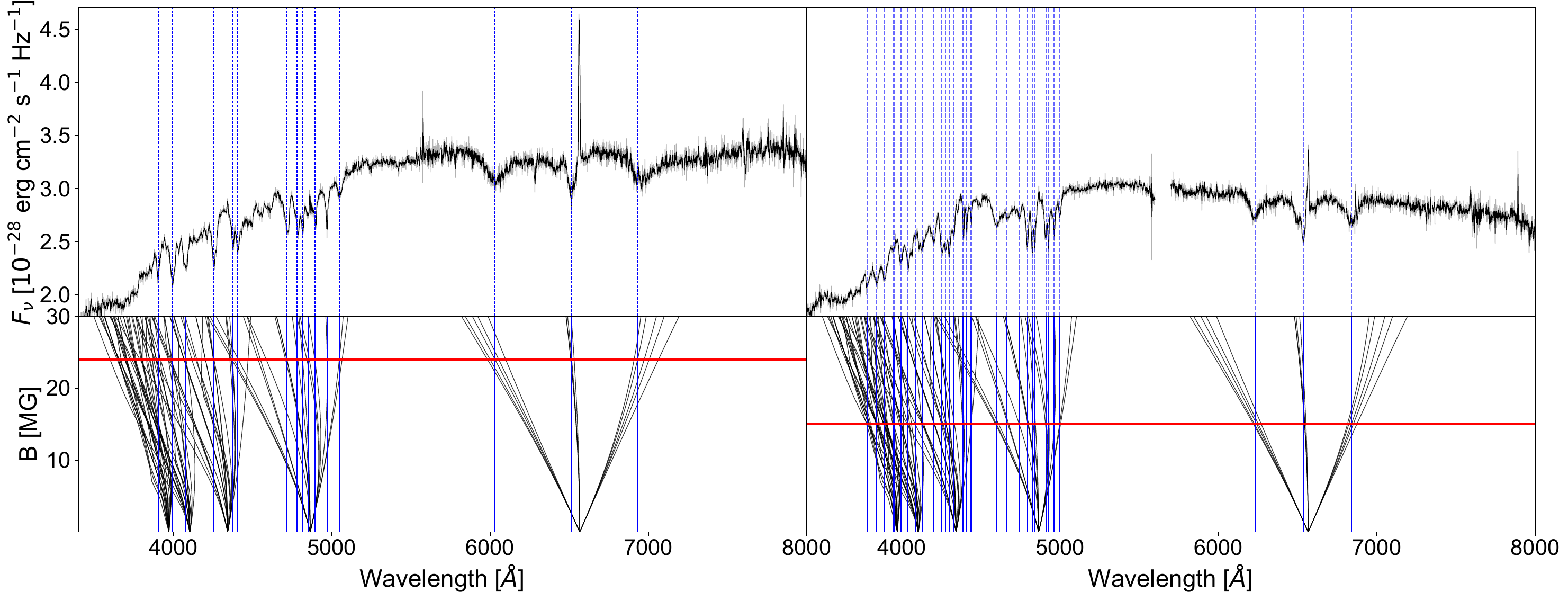}
	\caption{Phase-averaged optical spectra from Keck/LRIS of the two discovered systems with fits to the Zeeman split H absorption lines. Left and right panels show the spectra of WD\,J1820 and WD\,J1907, respectively. Both spectra show Zeeman split hydrogen Balmer absorption lines. The magnetic field strength is estimated by comparing the absorption line wavelengths with all bound-bound transitions at different field strengths \citep{kemic74}. We find the estimated field strength to be $B$\,=\,24 and 15\,MG for WD\,J1820 and WD,J1907, respectively. The spectra also show a narrow H$\alpha$ emission line around 6563\AA. The narrow feature implies the line is not formed at the white dwarf surface. Time-series spectroscopy reveals this line is consistent with being formed on the irradiated surface of a companion (see Figs.\,\ref{fg:trailed-spec-Halpha}--\ref{fg:Gaussian-periods}).}
	\label{fg:magnetic-fits-zeeman}
\end{figure*}

We obtained optical spectra using the Low-Resolution Imaging Spectrometer \citep[LRIS,][]{oke1995-LRIS} on the Keck I Telescope. We used the R600/4000 grism ($R$\,$\approx$\,$1100$) for the blue arm and the R400 grating ($R$\,$\approx$\,$1000$) for the red arm, covering a wavelength range of approximately 3200--10000\,\AA. A standard long-slit data reduction procedure was performed with the \texttt{Lpipe} pipeline\footnote{\url{http://www.astro.caltech.edu/~dperley/programs/lpipe.html}} \citep{perley2019}. During the first night, on May 23, 2023, we obtained two 15-minute exposures for each target; the spectra revealed two highly magnetized white dwarfs and a narrow H$\alpha$ emission around the non-magnetic rest wavelength that appeared to vary in both objects between the two exposures. We obtained an additional 8 consecutive 15-minute spectra for each target on June 16 and 17, 2023.
The additional spectra confirmed that the narrow emission line is variable both in strength and radial velocity on a period of about 100 minutes in both systems, with maximal Doppler shifts of respectively 200 and 350 km/s.
Fig.\,\ref{fg:magnetic-fits-zeeman} shows the combined, phase-averaged spectra for the two white dwarfs, while Fig.\,\ref{fg:trailed-spec-Halpha} shows trailed spectrograms for both systems, centered on the H-alpha component at 6562.8\AA.
The high radial velocity amplitude of the emission line, its periodic modulation, and its narrow profile, suggest that it is originating at the surface of the companion, rather than from the surface of the white dwarf. The most likely explanation for this is that the line is formed at the illuminated face of the companion, irradiated by one or more accretion hotspots on the white dwarf. However, no features from a companion are present in the spectra up to 1\,\micron, indicating that the companion must be either a very late type star, or sub stellar object such as a brown dwarf.

\subsection{Photometry}
We collected  photometric data on the two systems at all available wavelengths. For both objects, optical photometry is available from the Pan-STARRS PS1 survey \citep{ps1} and from Gaia DR3, which also provides the parallaxes and proper motions used in this study \citep{gaia2016,gaia2023}. In addition, UV photometry for WD\,J1907 is available from GALEX in the NUV \citep{martin2005}, and from the XMM Optical Monitor in the U, UVW1 and UVM2 filters at two different epochs, as it was obtained simultaneously with the EPIC camera X-ray data.

In the infrared, photometric data in the $J$, $H$ and $K_{\rm s}$ bands is available for WD\,J1820 from the UKIRT Infrared Deep Sky Survey catalog \citep[UKIDSS;][]{UKIDDS07}. Additionally, we obtained photometry in the $J$-band for both systems using the Wide-Field Infrared Camera \citep[WIRC;][]{wilson2003} on the 200-inch Hale Telescope at the Palomar Observatory. The full list of photometric observations used in this paper is presented in  Table~\ref{tab:photometry} and plotted in Fig.~\ref{fg:phot_fit_WD+BD}.

\begin{table}
	\centering
	\caption{Photometric measurements, Gaia parallaxes, and reddening measurements from the Bayestar19 dust map for the two systems. The optical and NUV data was used to obtain the white dwarf parameters, while the NIR data was used to constrain the nature of the sub-stellar companion and to detect cyclotron emission.}
	\label{tab:photometry}
	\begin{tabular}{l|lcc} 
		\hline \hline
		 & & WD J1820 & WD J1907\\
		\hline \hline
		\multirow{5}{*}{Pan-STARRS} 
                    & g & $20.13\pm0.02$ & $20.101\pm0.012$ \\
		          & r & $20.06\pm0.03$ & $20.05\pm0.02$ \\
                    & i & $19.96\pm0.04$ & $20.147\pm0.012$ \\
		          & z & $19.98\pm0.05$ & $20.14\pm0.11$\\
                    & y & $--$ & $19.82\pm 0.19$ \\
            \hline
            GALEX   & NUV & $--$ & $21.8\pm0.4$ \\
            \hline
            \multirow{5}{*}{XMM-OM} 
                & U obs. 1
                    & $--$ & $20.032\pm0.043$ \\
                    & U obs. 2 & $--$ & $20.761\pm0.104$ \\ [0.5ex]
                & UVW1 obs. 1 
                    & $--$ & $20.168\pm0.083$ \\
                    & UVW1 obs. 2 & $--$ & $20.68\pm0.081$ \\ [0.5ex]
                & UVM2 obs. 2 & $--$ & $21.39\pm0.29$ \\
            \hline
            WIRC & J & $20.19\pm0.10$ & $20.76\pm0.13$ \\
            \hline
            \multirow{3}{*}{UKIDSS} 
                & J & $18.83\pm0.11$ & $--$ \\
                & H & $18.89\pm0.15$ & $--$ \\
                & K$_{\rm s}$ & $20.1\pm0.4$ & $--$ \\
        \hline
            \multicolumn{2}{l}{Parallax} & $4.7\pm0.5$ & $4.2\pm0.4$ \\
            \multicolumn{2}{l}{E(B-V)} & $0.23\pm0.09$ & $0.19\pm0.08$ \\
            \hline \hline
	\end{tabular}
\end{table}

\subsection{Photometric variability}
\label{sec:photvar}
In the infrared, the two $J$-band observations (WIRC and UKIDSS) of WD\,J1820 show strong photometric variability, most likely due to cyclotron emission (see section~\ref{sec:cyclotron}). The SNR in the PS1 light-curve is too low to detect variability, but the excess in PS1-y for WD\,J1907 is likely caused by cyclotron emission as well.
The magnetic field in the two systems is not high enough for cyclotron to contribute much in the optical, and we expect most of the variability to happen in the near IR.
We analyzed the optical light curves of the two systems in g and r band in the Zwicky Transient Facility archive \citep[ZTF;][]{bellm19}. A Lomb-Scargle search reveals no strong indication of a periodic variability in either system. We cannot exclude a low-amplitude optical variability as both objects are rather faint, and the low signal-to-noise ZTF observations imply that only a modulation larger than about 10\% peak to peak can be detected.

In the UV, we only have data for WD\,J1907. The XMM-OM observations show a dimming of over half a magnitude between the two epochs in U and UVW1, while the UVM2 observation (only available for the second epoch) is in agreement with the only GALEX observation. As we explain below, the second (dimmer) epoch is in agreement with the expected UV flux from the white dwarf. The brightening could be due to a transient phenomenon, like a flare, or to periodic variability, as for example if a small hot spot is present on the surface of the white dwarf. Similarly to EF Eri \cite[see also section\,\ref{sec:pop}]{szkody2006,schwope2007}, the UV data during the bright phase can be modeled by a small, $\approx$20,000\,K hotspot; however, additional monitoring in the UV would be necessary to understand the source of variability.

\section{Analysis}
\label{sec:analysis}

\subsection{White dwarf parameters from photometry}
\label{sec:WDfitting}
To determine the parameters of the two white dwarfs, we performed an SED-fitting to their optical (and UV) photometry, where we expect the white dwarf flux to dominate, using non-magnetic DA spectral models. We made use of the available optical photometry from PS1 and of Gaia DR3 parallaxes. In addition, we used GALEX NUV photometry, available only for WD\,J1907. As the XMM UV photometry for WD\,J1907 shows strong variability, possibly caused by flares or a hot spot (see section \ref{sec:photvar}), we do not include it in the fit. In the case of WD\,J1907, the reddest PS1 filter, PS1-$y$, shows a strong excess which is most likely due to cyclotron emission (see below), so we exclude this data point from our fitting. 

For the fitting, we employed the 1-D atmosphere models for non-magnetic DA (hydrogen dominated) white dwarfs developed by \citet{tremblay2011}. To account for extinction, we applied reddening corrections to the synthetic spectra using the \citet{cardelli1989} extinction curves\footnote{The \citet{cardelli1989} extinction curves are available at \url{https://www.stsci.edu/hst/instrumentation/reference-data-for-calibration-and-tools/astronomical-catalogs/interstellar-extinction-curves}.}. 
From the corrected models, we computed synthetic photometry using the \texttt{pyphot} package\footnote{The Python-based photometry package, \texttt{pyphot}, is available at \url{https://mfouesneau.github.io/docs/pyphot/}.}.
For the fit, we used a Levenberg-Marquardt algorithm, and the free parameters were the effective temperature of the white dwarf, $T_{\rm{eff}}$, the ratio between the radius of the white dwarf, $R_{\rm{WD}}$, and its distance from Earth, $D$, as well as the interstellar reddening, $E(B-V)$. For the reddening, we imposed a Gaussian prior based on the Bayestar19 dust map \citep{green2019}. From the ratio $R_{\rm{WD}}/D$ obtained from the fit, we calculated the radius of the white dwarf using the Gaia parallax and parallax error. From the radius and effective temperature, we obtained a mass and cooling age using DA evolutionary models \citep{bedard2020}. The results are listed in Table\,\ref{tab:Phot_fitting_optical} and the best-fitting models are shown in the upper panels of Fig.\,\ref{fg:phot_fit_WD+BD}. We can see that the model fitted to the PS1 and GALEX photometry for WD\,J1907 is in agreement with the XMM-OM data during the second (dimmer) observation.

\subsection{White dwarf magnetic fields}

In Fig.\,\ref{fg:magnetic-fits-zeeman}, we identify the observed absorption features in both objects as Zeeman split Balmer components, comparing them to all bound-bound transitions at different field strengths \citep{kemic74}. From the splitting, we estimate the average magnetic field on the surface to be about 24\,MG for WD\,J1820 and 15\,MG for WD\,J1907; the magnetic field at the pole is likely to be higher. These field strengths in excess of 10\,MG indicate that the systems are likely polars. Additionally, we have no indication of the presence of an accretion disk in the form of broad or double-peaked emission lines or excess in the optical and UV. It would be interesting to see if the white dwarfs in these systems that have been evolving for Gyrs past the period bounce (see section\,\ref{sec:evol}) are still synchronized with the orbital period, as it is usually seen in polars.  Unfortunately, we have currently no evidence on the rotation period of the white dwarfs; however, if light curves in the IR are obtained, they would reveal the periodicity of the cyclotron emission from the white dwarfs and confirm their spin periods. We did perform a time-series analysis of the X-ray event files. Whilst tentative periods were identified on or near the orbital (H$\alpha$) period, the signal-to-noise was inadequate to draw out a confident detection of a period.

\subsection{Orbital parameters}
\label{sec:analysis-orbital}

\begin{figure}
	\centering

    \subfloat{\includegraphics[width=0.5\columnwidth]{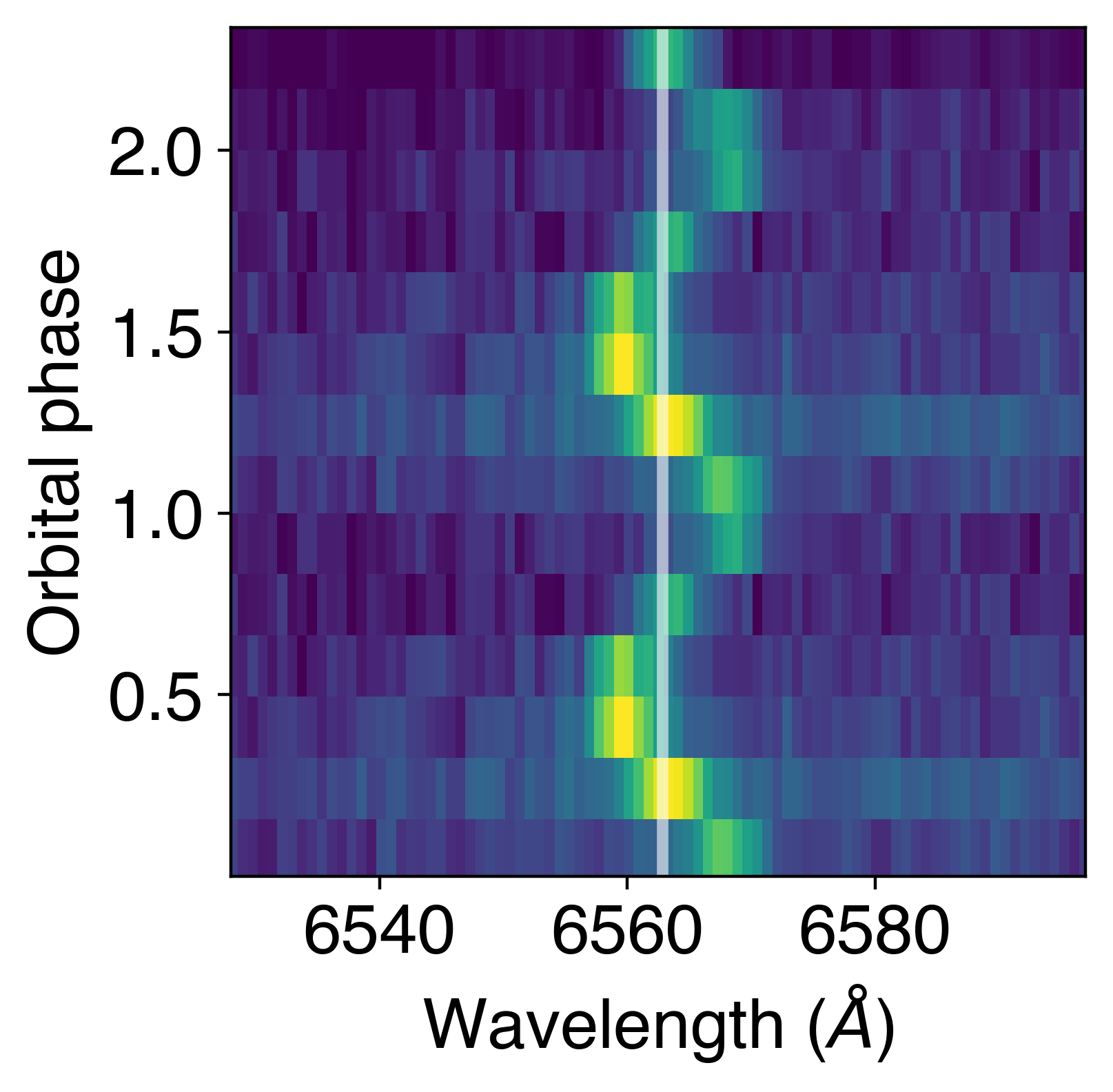}}
    \subfloat{\includegraphics[width=0.5\columnwidth]{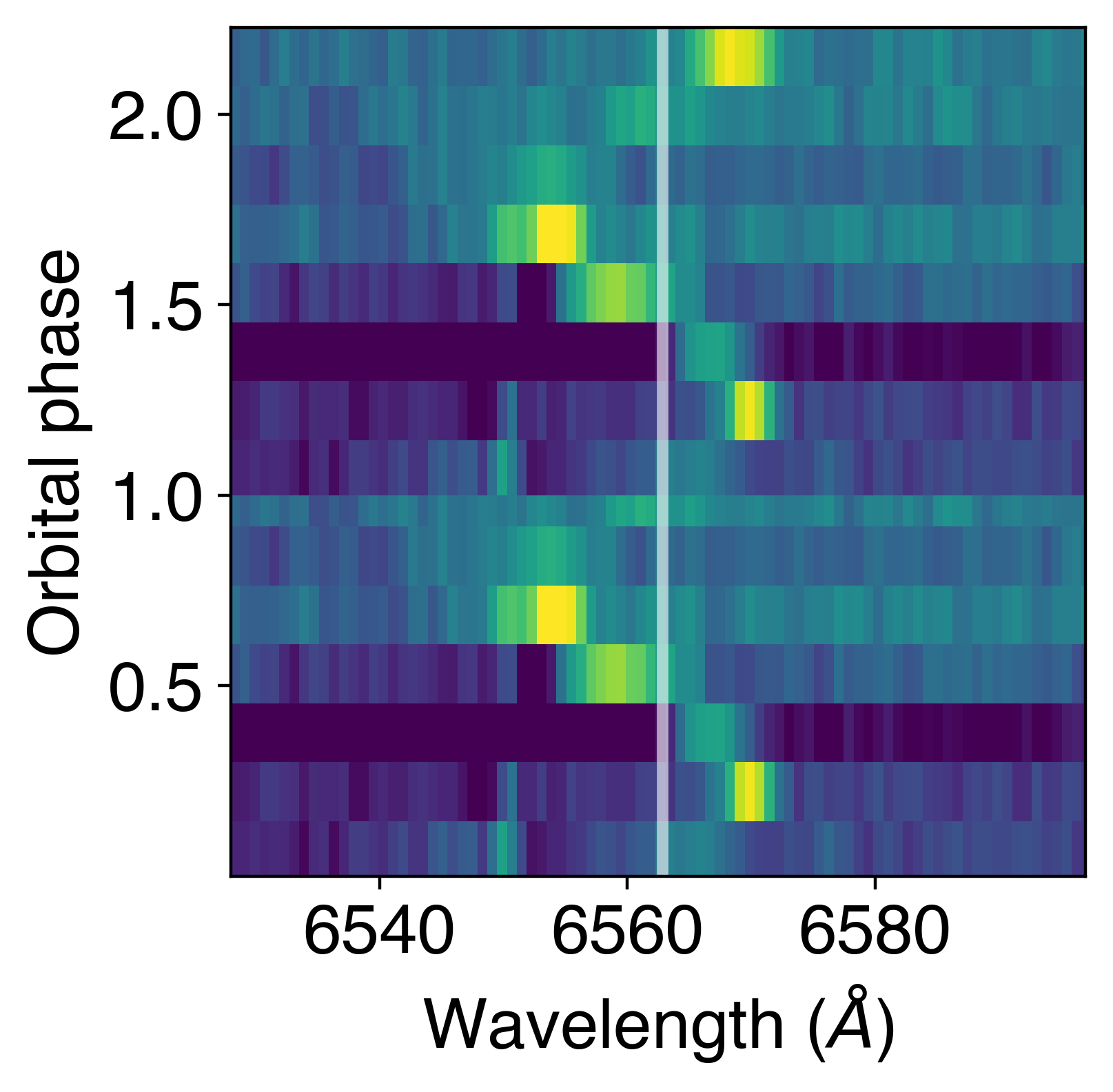}}
    \\
    \subfloat{\includegraphics[width=0.5\columnwidth]{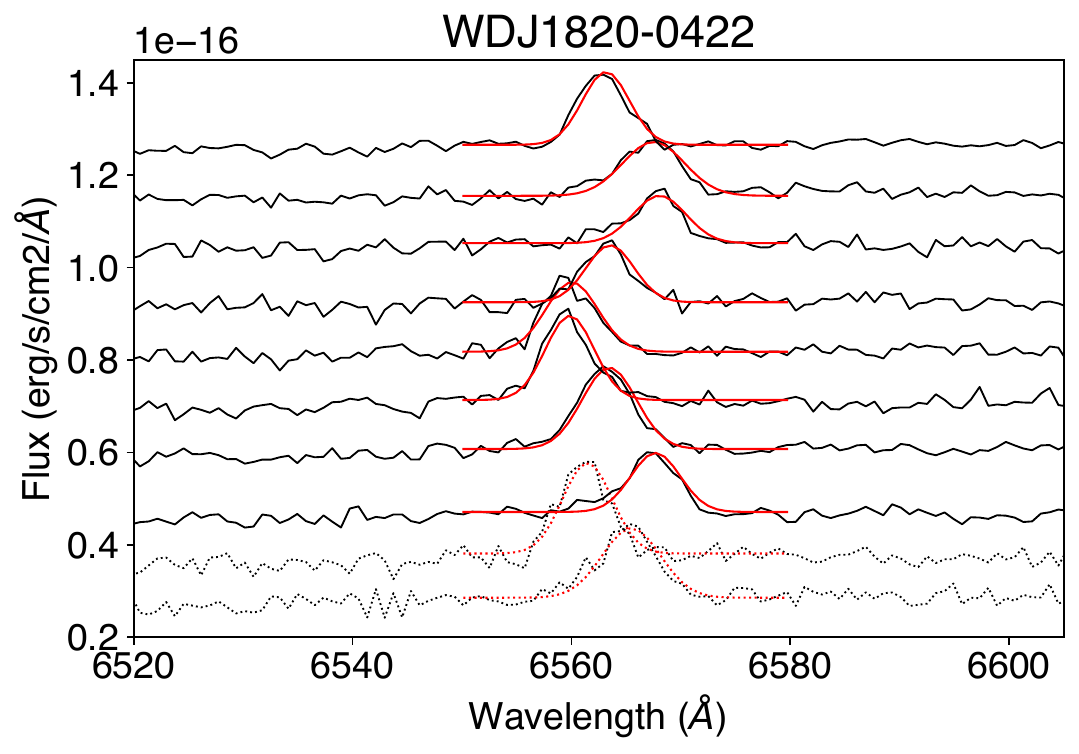}}
    \subfloat{\includegraphics[width=0.5\columnwidth]{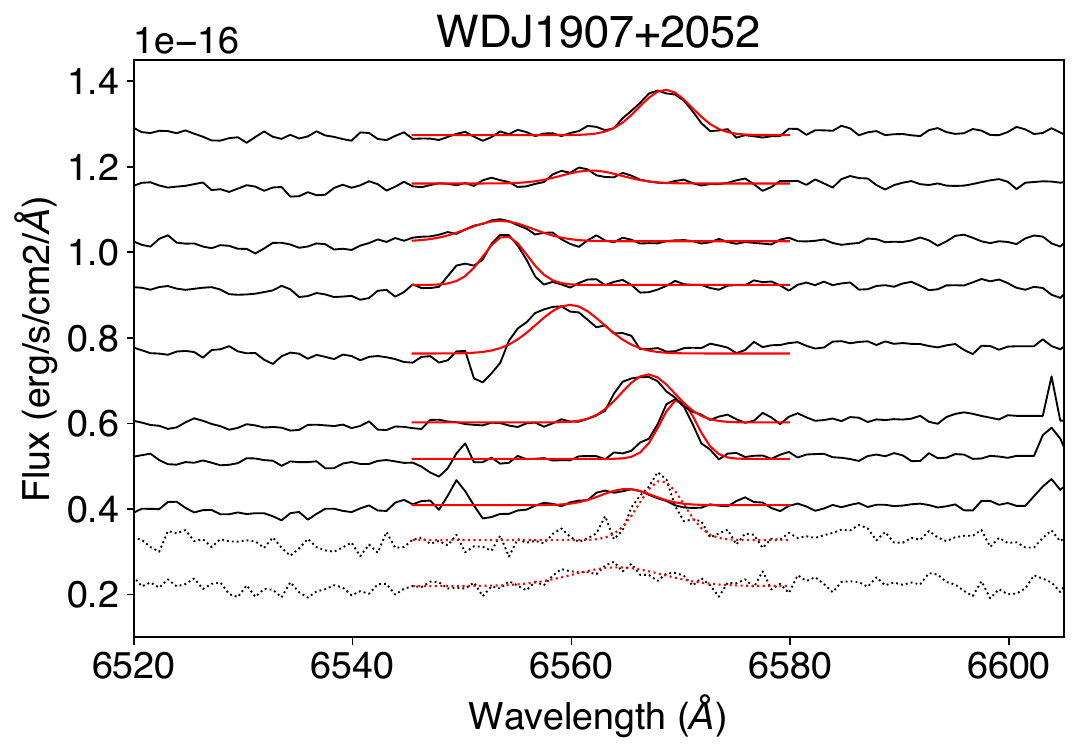}}
  
	\caption{{\bf Top}: Trailed spectrograms centered on H-alpha at 6562.8\AA\ for the two newly-discovered systems presented in this paper. The spectrograms have been phase-folded on periods of 95.3 and 103.8\,min, on the left and right, respectively. {\bf Bottom}: Gaussian fits to the H-alpha emission line for the eight 900-second Keck/LRIS spectra taken during the time series follow-up (solid), and the two 900-second consecutive spectra taken during the initial spectroscopic identification observations (dashed). The data are shown in black, whilst the Fitted Gaussians are shown in red. For clarity, the spectra and fits are given a vertical offset to separate them.}
	\label{fg:Halpha-fits}
        \label{fg:trailed-spec-Halpha}
\end{figure}

The periodic modulation in intensity, radial velocity amplitude, and small equivalent width of the H$\alpha$ emission line suggest that the line is created on the irradiated face of the sub-stellar companion, as seen in other low-state polars and period bouncers \citep{schmidt1996,breedt2012,kawka2021}. We can therefore study its modulation to extract the orbital parameters of the binary. We fit Gaussian profiles to the narrow H-alpha emission line detected in all ten spectra taken for both systems. In the fitting procedure, we allow for 4 free parameters: the central wavelength, amplitude, standard deviation and continuum level. The equivalent width does not show significant variations between observations in either system, as is expected from irradiation; we therefore performed a global fit to the emission line in all observations, keeping the equivalent width the same across all observations, and we found no significant change in the other fitted parameters.  The central wavelength gives a measurement of the radial velocity, the amplitude and standard deviation provide a measurement of the integrated line flux, and the continuum level allows to remove the contribution of the white dwarf continuum. The Gaussian fits to the time-series data are shown in Fig.\,\ref{fg:Halpha-fits}. The data is shown in black, and the Gaussian fits in red. The two initial identification spectra for each system are shown at the bottom of the plot with fainter lines. The eight consecutive follow-up spectra are shown above with emboldened lines. In the following we study the time dependence of the fitted Gaussian parameters.

\subsubsection{Orbital period}
\begin{figure}
	\centering

    \subfloat{\includegraphics[width=0.49\columnwidth]{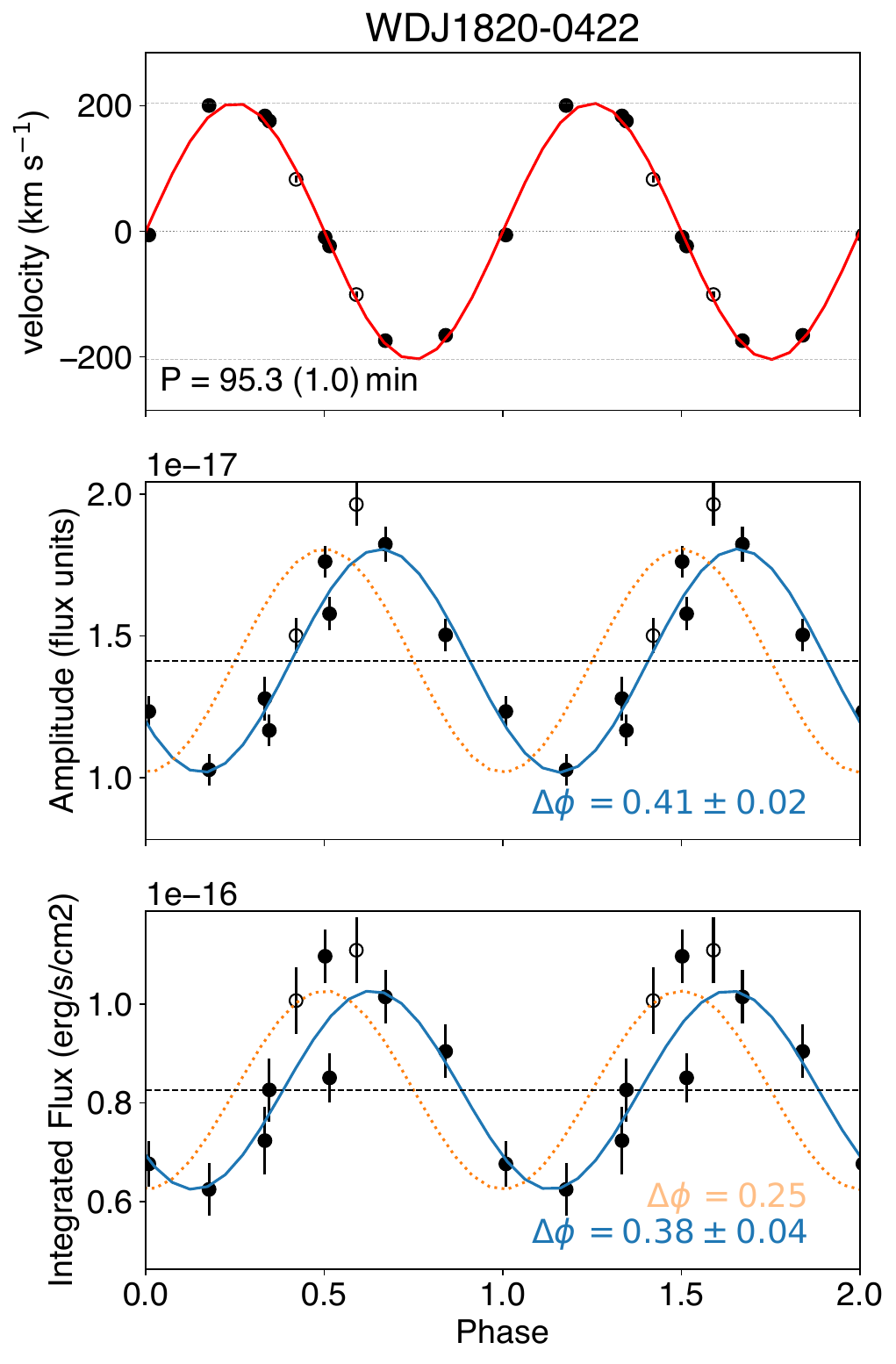}}
    \subfloat{\includegraphics[width=0.49\columnwidth]{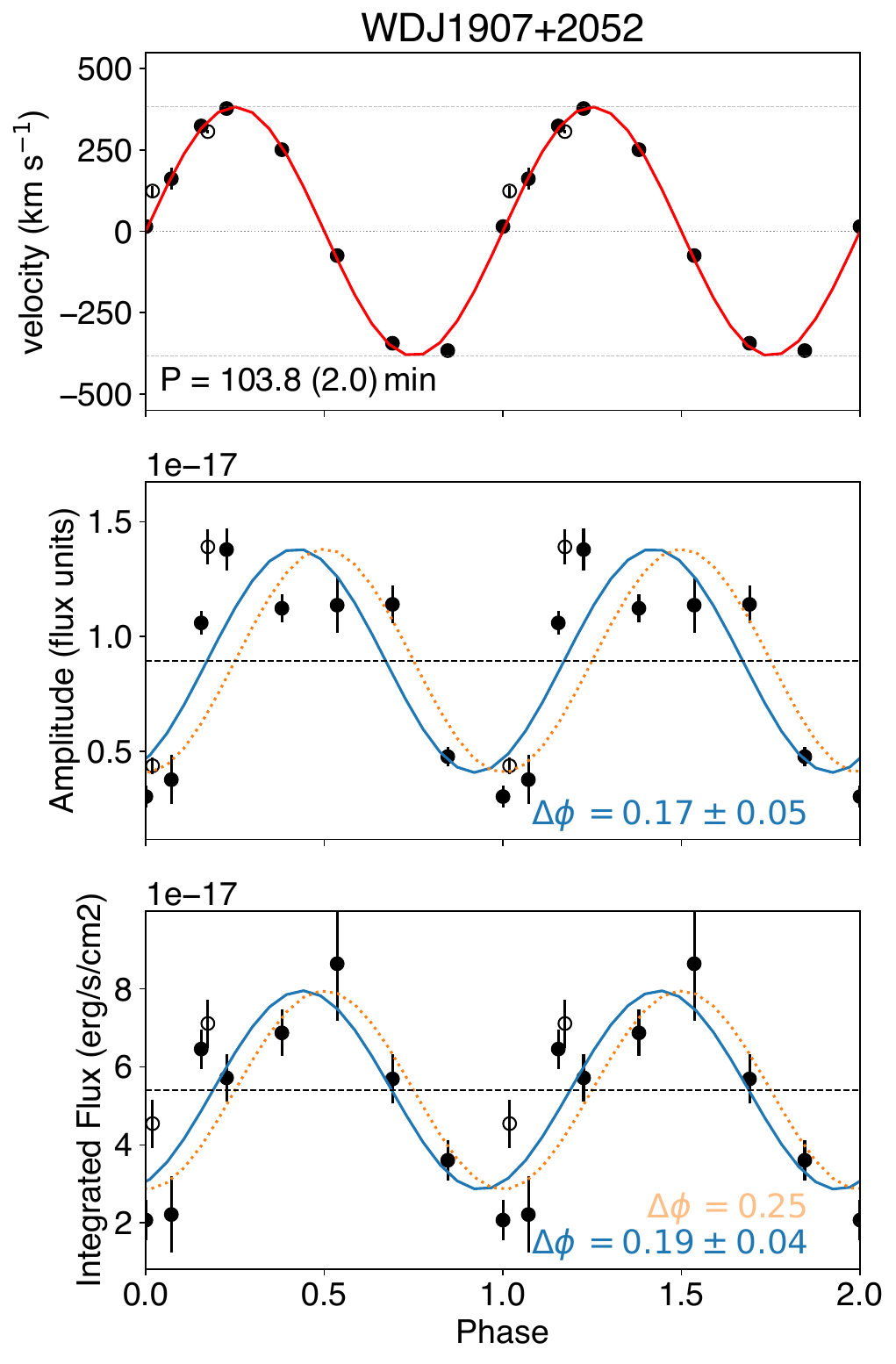}}
  
	\caption{Observed variations in the parameters of the H$\alpha$ emission line compared to our best-fitting sinusoidal models for WD\,J1820 (left) and WD\,J1907 (right). The measurements shown originate from 2 consecutive 15-min spectra taken in June 2023 (open circles) and 8 consecutive 15-min spectra taken in July 2023 (filled circles) for each object. \textbf{Top}: Radial velocity of the emission line. The red curve shows the best-fitting 4-parameter sinusoid, fitted only to the 8 consecutive spectra (filled circles). The open circles are omitted from the fit and have been shifted in phase by a fixed amount for each system to match the RV curve. We find a sinusoidal period of $P_{\rm orb}=95.3\pm1.0$\,min and of $P_{\rm orb}=103.8\pm2.0$\,min for WD\,J1820 (left) and for WD\,J1907 (right), respectively. \textbf{Middle}: Amplitude of the emission line  above the local continuum. In blue we show a 3-parameter sinusoidal fit, where the period is fixed at the value obtained from the radial velocities (top panel), but the amplitude, phase and normalisation are left free. Again, only the eight filled circles are included in this fit. The fitted offset in phase with respect to the top panel, $\delta\phi$, is shown in the panels in blue. The orange dotted line shows the same curve, but at a phase offset of a quarter period. \textbf{Bottom}: Same as middle panel, but with the integrated flux over the H-alpha emission line instead of the amplitude.}
	\label{fg:Gaussian-periods}
\end{figure}

We find the orbital periods of the two systems by fitting a sinusoid to the radial velocities of the H-alpha emission obtained from the eight consecutive spectra for each system. We find the best-fit orbital periods for the two systems to be $P_{\rm orb}=95\pm1$ and $104\pm2$\,min, for WD\,J1820 and WD\,J1907, respectively. The radial velocities and the best fit sinusoids are shown in Fig.\,\ref{fg:Gaussian-periods}. Although we do not use the initial identification spectra in the fit, as they are separated by hundreds of periods, we show their radial velocity measurements in open circles, phase-folded at the best-fit periods.

\subsubsection{Phase offset}
In Fig.\,\ref{fg:Gaussian-periods}, the middle panels show the time-dependent Gaussian amplitude of the H$\alpha$ emission line. In blue, we show a sinusoidal fit to this data, in this instance fitting just three parameters: the amplitude, phase, and mean $y$-value. We fix the period of the sinusoid to that measured from the radial velocities. The idea behind this test is to constrain the phase offset between the radial velocity and the amplitude. The expectation is that the phase of the amplitude should be a quarter phase behind the radial velocity curve. Assuming the system to be tidally locked, the irradiated surface of the companion brown dwarf, where the H$\alpha$ line is expected to be formed, should be maximally visible when the brown dwarf is behind the white dwarf. At this point, which we call phase $\phi=0.5$, the line-of-sight velocity of the brown dwarf is zero. At phase $\phi=0$, the irradiated surface of the brown dwarf should be facing away from the observer, at which point the measured H-alpha line intensity ought to be at a minimum. In the middle panel, we indicate in blue text the phase offset between the fit to the radial velocities and the fit to the amplitude. We find the best-fit phase offset to be $\Delta\phi=0.41\pm 0.05$ and $0.17\pm0.05$, for WD\,J1820 and WD\,J1907, respectively. We also show the expected curve with an imposed phase offset of $\Delta\phi=0.25$ (dotted orange line). We find that the phase offset in WD\,J1907 is consistent with the expected quarter phase within the errors. On the other hand, WD\,J1820, appears to show a larger phase offset, in which the peak amplitude of the H$\alpha$ line appears delayed by an additional 0.15 in phase with respect to the expected quarter value. In the bottom panel we show the curve of integrated flux, rather than just the peak amplitude. We perform the same phase offset analysis and recover results that are consistent with those in the middle panel. Phase offsets have been observed in CVs with higher accretion rates \citep{beuermann1990}, thought to originate from an illuminating source offset from the center of mass. Such a geometry could arise if the source of irradiation is, for example, a hot stand-off shock sufficiently high above the white dwarf surface. Alternatively, an offset could be due to the accretion stream shielding part of the brown dwarf from the X-ray emission that is causing the irradiation.

\subsubsection{Inclination}
\label{sec:analysis-orbital-inclination}
The inclination $i$ of the system with respect to the line of sight affects the maximum and minimum amplitude in the H$\alpha$ line. Under the assumption that the night side of the brown dwarf produces no emission, in an edge-on system ($i=90^{\circ}$) the H$\alpha$ emission should go to zero when the night side of the brown dwarf is facing our direction, i.e at phase $\phi=0$. We can therefore derive an estimation of the inclination of the binary as the inverse sine of the ratio between the maximum and mean flux in the line \citep[see for example][]{beuermann1990,schwope2010}. In case some emission is produced beyond the irradiated hemisphere, i.e. on the night side of the companion, then this estimate becomes a lower limit.
Performing this calculation on the measured mean flux and amplitude for the peak H-alpha emission (middle panel of Fig.\,\ref{fg:Gaussian-periods}) yields a lower limit on the inclination of $i\geq(14\pm2)^{\circ}$ and $i\geq(28\pm10)^{\circ}$, for WD\,J1820 and WD\,J1907, respectively. For the integrated fluxes (lower panel), we recover consistent lower limits of $i\geq(16\pm4)^{\circ}$ and $i\geq(32\pm6)^{\circ}$.

\subsection{Constraints on the donors}
\label{sec:donors}
\begin{figure*}
	\centering
        \includegraphics[width=0.48\textwidth]{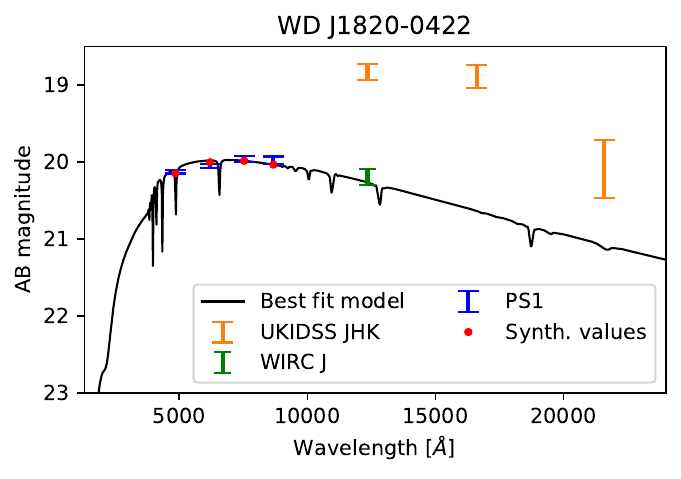}\includegraphics[width=0.51\textwidth]{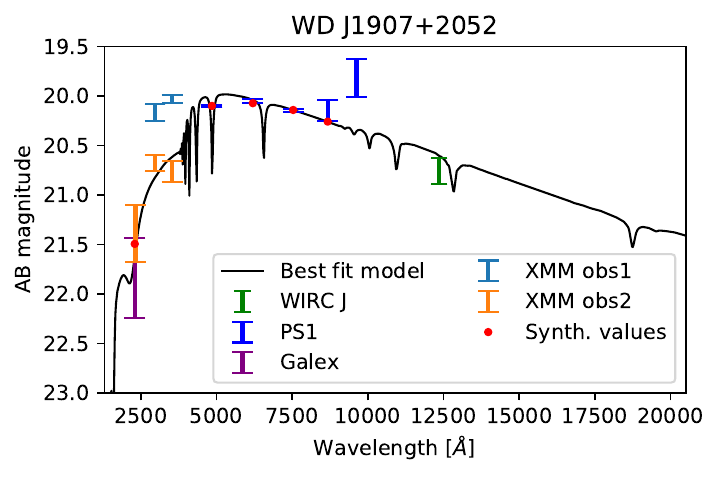}
	\includegraphics[width=0.48\textwidth]{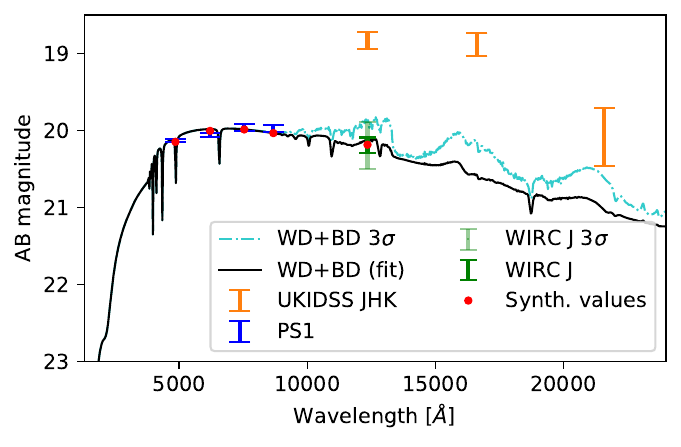}
 	\includegraphics[width=0.51\textwidth]{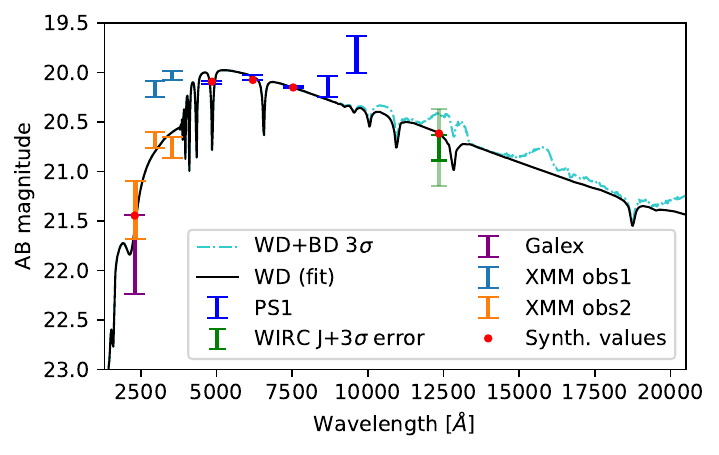}

	\caption{Photometric data available for the two systems: {\bf left} WD\,J1820, {\bf right} WD\,J1907 (see Table\,\ref{tab:photometry}). For WD\,J1820, the available data includes PS1 photometry in $g$, $r$, $i$ and $z$ (in blue) and UKIDSS in J, H and K$_s$ (in yellow); in addition, we obtained WIRC $J$-band (green). For WD\,J1907 we show PS1; $g$, $r$, $i$, $z$ and $y$ (in blue), GALEX NUV (in purple), XMM-OM U and UVW1 at two different epochs (light blue and yellow) and UVM2 only in the second epoch (yellow), plus our WIRC $J$-band (green). {\bf Upper plots}: Best-fitting white dwarf models to the optical and UV data as solid black lines (see section\,\ref{sec:WDfitting} and Table\,\ref{tab:Phot_fitting_optical}). The red dots indicate the synthetic magnitudes extracted from the best-fitting models. For WD\,J1820, only PS1 data was included in the fitting, while for WD\,J1907, PS1 and GALEX. {\bf Lower plots}: Both systems show evidence of cyclotron emission in the infrared; however, the two WIRC $J$-band observations were likely taken close to the minimum in the cyclotron emission, and can be used to constrain the contribution from the donor. For WD\,J1820, we show the fit to the PS1 and WIRC data using a combined white dwarf plus brown dwarf model (sold black line and red markers); additionally, we plot the combined WD+BD model with the brightest BD allowed by the 3$\sigma$ error in the WIRC $J$-band magnitude (cyan). For WD\,J1907, the WIRC $J$-band data does not allow any excess with respect to the WD continuum; we thus cannot attempt a combined fit. We show only the best-fit white dwarf model in black (we now include the $J$-band in the fitting and we obtain the same best-fit model as in the upper plot) and the combined WD+BD model with the brightest BD allowed by the 3$\sigma$ error in the WIRC $J$-magnitude (cyan).}
	\label{fg:phot_fit_WD+BD}
\end{figure*}

To understand the nature of the companions, we obtained near-infrared photometry in the $J$-band using the Wide-Field Infrared Camera \citep[WIRC,][]{wilson2003} on the 200-inch Hale Telescope at the Palomar Observatory (see Table~\ref{tab:photometry} and Fig.~\ref{fg:phot_fit_WD+BD}). For WD\,J1820, photometric data in the $J$, $H$ and $K_{\rm s}$ bands is also available from the UKIRT Infrared Deep Sky Survey (UKIDSS) catalog. A comparison between the WIRC and UKIDSS photometry shows a strong variation of at least a magnitude in $J$ band, most likely due to cyclotron emission. The WIRC data in both objects is consistent with the white dwarf continuum, as inferred from optical and UV data (see sec.\,\ref{sec:WDfitting}), which tells us that the donor stars contribute very little in the J band. We here analyze the SED of the two objects including possible contributions from brown dwarf companions. For the brown dwarf models,  we use the ATMO2020 synthetic atmospheres \citep{ATMO2020} with a surface gravity of $\log g=5$ (we find that using different values of $\log g$ does not change our results, as our fits are not very sensitive to the surface gravity).

For WD\,J1907, the $J$-band data rules out almost any contribution from the companion. If we include the $J$-band data in the same SED fitting performed in section\,\ref{sec:WDfitting} (i.e, assuming that only the white dwarf contributes significant flux) we obtain identical parameters for the white dwarf as from the optical and UV data alone (Table\,\ref{tab:Phot_fitting_optical}). In the lower right panel of Fig.\,\ref{fg:phot_fit_WD+BD}, we show the best-fit white dwarf model (in black); additionally, we plot the sum of the white dwarf model plus the hottest brown dwarf that would be still consistent with the J band at its 3-$\sigma$ brightest limit (cyan line). At the distance of the white dwarf, we find that the hottest brown dwarf allowed by the data, assuming a radius of about 1 Jupiter radius, has a temperature of 1000\,K. We stress that this is not a fit but rather an upper limit. It is likely that the brown dwarf is actually much fainter.

The $J$-band flux in WD\,J1820 is consistent with the continuum of the best-fit white dwarf model to the optical data, but it allows also a small excess within one sigma. We therefore perform SED fitting with combined synthetic spectra from a white dwarf and a brown dwarf at the same distance.
The temperatures of the two objects, the reddening and the ratio $R_{\rm{WD}}/D$ are left as free parameters (imposing a prior on the reddening as done in the other fits), while we impose the radius of the brown dwarf to be 1 Jupiter radius. The resulting parameters are listed in Table~\ref{tab:Phot_fitting_WD+BD}; the white dwarf parameters are consistent with the ones found in section\,\ref{sec:WDfitting}, while we find that the best-fitting temperature for the brown dwarf is $1100\pm200$\,K, or a spectral type T5 \citep{faherty2016}. As there could still be some contribution from cyclotron emission in our measured $J$-band photometry, this is still an upper limit on the brown dwarf temperature.  In the lower left panel of Fig.~\ref{fg:phot_fit_WD+BD}, we show the best-fit combined model in cyan, and just the synthetic white dwarf spectrum in black. Also, we show the brightest brown dwarf allowed within the 3$\sigma$ limit of the $J$-band data, which has a temperature of 1,300\,K.

\begin{table}
	\centering
	\caption{Physical parameters of the two white dwarfs from fitting the optical (and UV) SEDs. Additionally, for WD\,J1820, we list the result of fitting the optical and IR data with combined white dwarf plus brown dwarf models.}
	\label{tab:Phot_fitting_optical}
        \label{tab:Phot_fitting_WD+BD}
	\begin{tabular}{lcc} 
		\hline
            \hline
		 & WD J1820 & WD J1907\\
		\hline
            \hline
		$T_{\rm{eff}}$ [K] & $9\,200\pm2\,400$& $11\,600\pm2\,100$ \\
		$R_{\rm{WD}}/D$ [km/pc]  & $35.7\pm6.2$ & $26.7\pm1.5$ \\
            E(B-V) & $0.19\pm0.17$ &  $0.19\pm0.08$ \\
		$R_{\rm{WD}}$ [km] & $7\,600\pm1\,500$ & $6\,200\pm600$\\
            $M_{\rm{WD}}$ [M$_\odot$] & $0.75^{+0.20}_{-0.18}$ & $0.92^{+0.08}_{-0.07}$ \\
		\hline
        $T_{\rm{eff}}$ WD [K] & $9\,300\pm2\,400$ \\
            $T_{\rm{eff}}$ BD [K] & $1\,100\pm200$ \\
		$R_{\rm{WD}}/D$ [km/pc]  & $35.2\pm6.2$ \\
            E(B-V) & $0.19\pm0.16$ \\
		$R_{\rm{WD}}$ [km] & $7\,500\pm1\,500$ \\
            $M_{\rm{WD}}$ [M$_\odot$] & $0.76^{+0.19}_{-0.18}$ \\      
		\hline
            \hline
	\end{tabular}
\end{table}

\subsection{X-ray analysis}
\label{sec:results-xray-fits}

\subsubsection{Spectroscopic fits}
The X-ray spectral analysis was performed using v4.1.1 of the Bayesian X-ray Analysis \citep[\texttt{BXA};][]{buchner2016-bxa} which connects the nested sampling algorithm, \texttt{UltraNest} \citep{buchner2019-UltraNest} with an X-ray spectral fitting environment. For the spectral fitting environment, we adopt v12.13.1 of \texttt{XSPEC} \citep{arnaud1996XSPEC} operated by its Python interface, \texttt{PyXSPEC}.
We fit the background-subtracted EPIC PN and MOS spectra simultaneously, using the C-statistic (\texttt{cstat}) within \texttt{XSPEC}. The spectra were first binned using the optimal binning algorithm of \citet{kaastra2016-optimal-binning}. We fit the EPIC spectra with a single isothermal plasma model (\texttt{apec}) 
and we include an absorption component (\texttt{tbabs}) to account for the Galactic nH absorption.
For the Galactic absorption toward the targets, we employ the extinction estimate from the Bayestar dust map (Table\,\ref{tab:photometry}) and convert it to hydrogen column density using the relation presented in \citet{guver2009}. We obtain N$_{\rm{H}}=(1.6\pm0.6)\times 10^{21}$\,cm$^{-2}$ and $(1.3\pm0.6)\times 10^{21}$\,cm$^{-2}$ for WD\,J1820 and WD\,J1907, respectively. We impose a uniform prior on the plasma temperature and a logarithmic uniform prior on the normalisation.

\begin{figure}
	\centering
	\includegraphics[width=0.495\columnwidth]{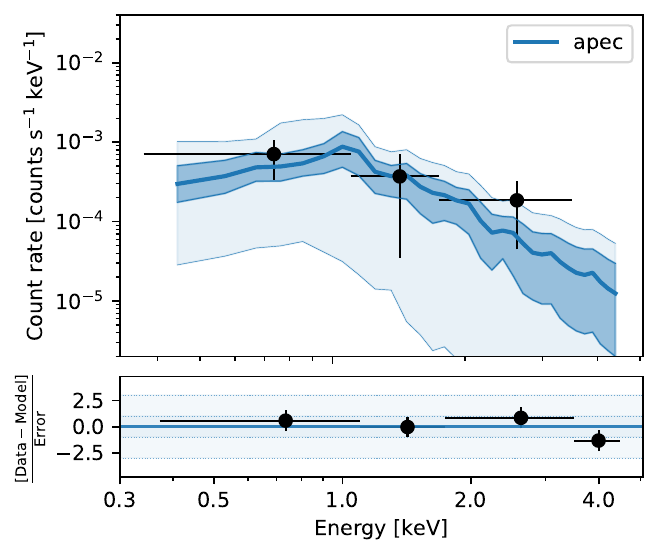} 
	\includegraphics[width=0.495\columnwidth]{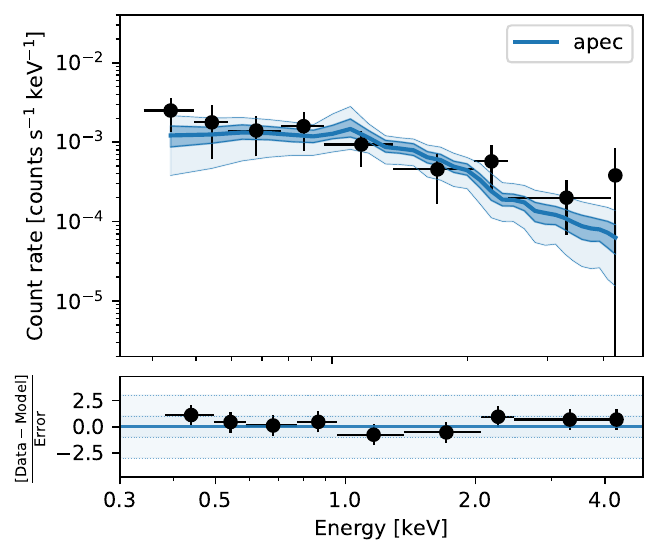}
    \includegraphics[width=0.495\columnwidth]{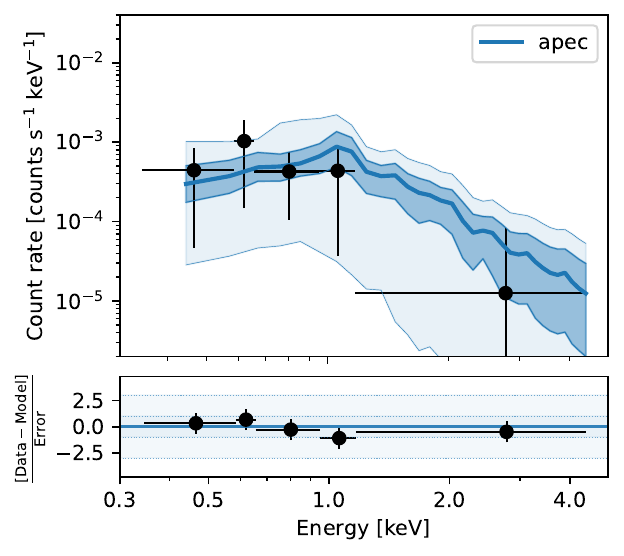} 
    \includegraphics[width=0.495\columnwidth]{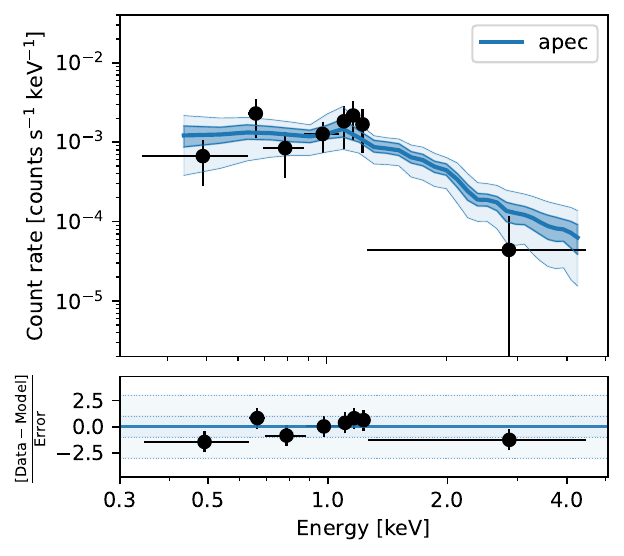}
	\caption{X-ray spectral fitting with BXA for WD\,J1820 (left) and WD\,J1907 (right), using data from both XMM EPIC detectors: PN (top) and MOS (M1 and M2 combined, bottom).  For each source, we fit the PN and MOS spectra simultaneously with an absorbed, isothermal plasma (APEC; blue) model. The shaded region corresponds to bands of uncertainty at 1 and 3\,$\sigma$ (darker and lighter respectively). The fitted spectra were binned using the optimal binning of \citet{kaastra2016-optimal-binning}. The spectra (black data points) have been re-binned to bins of higher significance for plotting purposes only.}
	\label{fg:bxa-fitting}
\end{figure}

\begin{table}
\centering
\caption{X-ray spectral parameters from the BXA fit for WD\,J1820 and WD\,J1907. 
We employ a single-temperature plasma model (APEC) with an additional absorption component.}
\label{tab:bxa-bb-apec}
\begin{tabular}{l|lllll}
\hline
\hline

 & & \multicolumn{2}{c}{68\%} & \multicolumn{2}{c|}{90\%} \\ 
 & best-fit & low & high & low & \multicolumn{1}{l|}{high} \\ [0.5ex]
										
\hline					
\texttt{tbabs*apec} & \multicolumn{5}{c}{\hspace{10pt}WD\,J1820}\\
\hline		
kT$_{\rm apec}$\,[keV] &	\textbf{1.96} &	1.14 &	3.64 &	0.76 &	4.45 \\
norm$_{\rm apec}$\,[10$^{-6}$]  &	\textbf{1.90} &	0.91 &	3.23 &	0.48 &	3.97 \\
nH [10$^{20}$\,cm$^{-2}$] &	\textbf{12.80} &	7.21 &	18.70 &	3.44 &	22.00 \\

\hline
\texttt{tbabs*apec} & \multicolumn{5}{c}{\hspace{10pt}WD\,J1907}\\
\hline

kT$_{\rm apec}$\,[keV] &	\textbf{3.85} &	2.49 &	6.75 &	2.26 &	7.85 \\
norm$_{\rm apec}$\,[10$^{-6}$]  &	\textbf{7.43} &	6.09 &	8.89 &	5.71 &	9.25 \\
nH [10$^{20}$\,cm$^{-2}$] &	\textbf{5.36} &	0.79 &	10.30 &	0.01 &	11.90 \\

\hline
\hline

\end{tabular}
\end{table}

\begin{table}
	\centering
	\caption{X-ray properties of WD\,J1820 and WD\,J1907 derived from BXA fit to the data as part of this study. The measured X-ray flux and all derived X-ray parameters correspond to the energy band 0.25--10.0\,keV. The 68\% and 90\% confidence limits on these parameters, and those for a soft band (0.25--2.0\,keV), can be found in the Appendix Tables\,\ref{tab:app-bxa-1820}\,\&\,\ref{tab:app-bxa-1907}. For comparison, we list the lower limit on the cyclotron flux that we derive from the infrared photometry. We also include the bolometric X-ray luminosity, $L_{\rm X, bol}$, computed by integrating the best-fit, unabsorbed model across the 0.0001--100\,keV energy band.}
	\label{tab:bxa-bb-apec-Mdot}
	\begin{tabular}{lrr} 
    \hline
		\hline
		 &  WD\,J1820 & WD\,J1907\\
		\hline

Flux [$10^{-15}$\,erg/s/cm$^2$] &
${2.4^{+1.6}_{-1.1}}$  & ${11.2^{+2.8}_{-3.4}}$\vspace{5pt}\\

Distance [pc]		& $213 \pm 22$		    & $241 \pm 23$\vspace{5pt}   \\
Luminosity [$10^{28}$\,erg/s]  & $1.2_{-0.6}^{+0.8}$	& ${11.4^{+3.0}_{-3.4}}$\vspace{5pt} \\

\Mwd\ [\Msun]		    & $0.76 \pm 0.2$        & $0.92 \pm 0.1$  \vspace{5pt}\\
$R_{\odot}$ [$10^3$\,km]		& $7.5 \pm 1.5$       & $6.2 \pm 0.7$\vspace{5pt}\\
$\dot{M}_{\rm X}$ [$10^{11}$\,g/s]& ${1.8_{-0.8}^{+1.2}}$ & ${11.6_{-3.1}^{+3.5}}$\vspace{5pt}\\

$\dot{M}_{\rm X}$ [$10^{-14}$\,\Msun/yr] & ${0.3_{-0.1}^{+0.2}}$ & ${1.8_{-0.5}^{+0.6}}$\vspace{5pt}\\

$kT$ [keV] & ${2.0_{-0.8}^{+1.7}}$ & ${3.9_{-1.4}^{+2.9}}$\vspace{5pt}\\

$L_{\rm X, bol}$ [$10^{28}$\,erg/s]  & $2.6_{-0.9}^{+1.2}$	& ${10.0^{+2.7}_{-2.2}}$\vspace{5pt} \\
\hline
F$_{\rm{cycl}}$ [10$^{-15}$ erg/s/cm$^2$] & $\gtrsim$50 & -- \vspace{5pt}\\
		\hline\hline
	\end{tabular}
\end{table}

\subsubsection{Plasma temperature}
Table\,\ref{tab:bxa-bb-apec} shows the best-fit parameters from the spectroscopic fits of the two systems. For WD\,J1907, we find an APEC plasma temperature of $kT = 3.9_{-1.4}^{+2.9}$\,keV, while for  
 WD\,J1820, we find $kT = 2.0_{-0.8}^{+1.7}$\,keV.
The APEC plasma temperature for WD\,J1907 is a factor two higher than that for WD\,J1820. The measured X-ray luminosity is higher by the same factor, which may suggest some correlation between the accretion rate and plasma temperature. The measured temperatures are broadly consistent consistent with the plasma temperature measured for the only other magnetic period-bounce CV observed with XMM; SDSS\,J1514, for which \citet{munoz-giraldo2023} derived a temperature (and 90\% confidence limits) of  $kT=3.75^{+5.92}_{-2.55}$\,keV. This is also broadly consistent with typical temperatures measured for of polars in low-states which tend to be around a few keV \citep{ramsay2004}.
Recent observations with the \textit{Chandra X-ray Observatory} \citep{pence2001Chandra} of the planet-debris accreting white dwarf, G29--38 \citep{cunningham2022}, yielded a measured plasma temperature of $kT=0.5\pm0.2$\,keV, providing  evidence in support of the bombardment accretion scenario proposed by \citet{kuijpers1982}. In this model, at sufficiently low accretion rates, the accretion column is unable to support a stand-off shock, instead releasing the gravitational potential energy of the infalling material directly into the atmosphere of the white dwarf. This scenario was predicted to yield significantly lower plasma temperatures compared to those typically measured from stand-off shocks above the surface of accreting white dwarfs. Given that the systems presented in this work  are among the lowest accretion rate CVs known, we consider this model in the context of these newly discovered systems. The temperature of the emitting plasma in the bombardment solution model was defined by \citep[][see their Equation 9]{kuijpers1982} to depend simply on the accretor mass and radius. Adopting their prescription, the predicted plasma temperature for WD\,J1820 and WD\,J1907 is $kT$\,=\,$0.87$ and 1.4\,keV, respectively. The 90\% confidence limit on the X-ray plasma temperature (see Table\,\ref{tab:bxa-bb-apec-Mdot}) for WD\,J1820 is consistent with the prediction from the bombardment solution, but the APEC temperature for WD\,J1907 is higher. We note that the latter does have an accretion rate 1 dex larger than the former, which could give an indication of a transition from shock to bombardment. However, as we currently do not know how much of the accretion-induced luminosity is emitted at other wavelengths, it is too early to draw a conclusion. A detailed X-ray study of the growing sample of low accretion rate white dwarfs, bridging between period bouncers and white dwarfs accreting from planetary debris, would provide the possibility to investigate this transition.

\subsubsection{Accretion rates}
\label{sec:accrate}
The X-ray luminosity is computed using the measured X-ray flux, 
and the distance determined from the \textit{Gaia} parallax. The accretion rate is estimated following the prescription of \citet{patterson85}, such that 
\begin{equation}
 \dot{M_{\rm X}} = \frac{2}{A}L_{\rm X} \frac{R_{\rm WD}}{GM_{\rm WD}}\, ,
 \label{eq:Mdot-Xray}
\end{equation}
where the factor two accounts for half of the emitted photons being directed back towards the star \citep{kylafis1982} and the constant $A$ encapsulates the fraction of the accretion-induced flux emitted outside of the XMM passband. In our derivation of the instantaneous accretion rate, we adopt the limiting case that no additional flux is carried at wavelengths outside of the XMM passband. In so doing, the quoted accretion rates are lower limits on the real accretion rate. We revisit the possibility of accretion-induced luminosity emitted at other wavelengths in Section\,\ref{sec:cyclotron}.
Table\,\ref{tab:bxa-bb-apec-Mdot} shows the higher order X-ray properties derived from the measured X-ray flux. All the parameters in the table are derived from the X-ray flux in the broad band 0.25--10.0\,keV. We include additional bands and confidence limits in the Appendix Tables\,\ref{tab:app-bxa-1820}\,\&\,\ref{tab:app-bxa-1907}. We find WD\,J1907 to be 1 dex higher in luminosity, but just a factor 4 in accretion rate, owing to the higher mass and smaller radius of the white dwarf.

\section{Discussion}
\label{sec:discussion} 

The strong magnetic fields, low accretion rates, long periods and the sub-stellar nature of the donors characterise these two systems as strong candidates to be magnetic period bouncers. As we discuss in section\,\ref{sec:pop}, there are very few known candidates, of which only 3 have confirmed sub-stellar donors, only one of which has a period above 90 minutes. Therefore, these newly discovered bouncers provide a great opportunity to study the evolution of CVs post period bounce. In this section, we discuss how our observations can constrain the properties of the systems.

\subsection{Orbital dynamics}
In Section\,\ref{sec:analysis-orbital} we derived orbital parameters from the time-series spectroscopy. Primarily, we derived a period and amplitude from the radial velocity of the H-alpha line, as well as a lower limit on the inclination from the H-alpha amplitude variation, without making assumptions on the accretion mechanism. Accretion in these systems could be due to the donors filling their Roche lobes. Alternatively, \citet{schreiber2023} suggested that accretion in some magnetic period bouncers might be driven by weak winds from the donors, although a significant mass loss through winds has never been detected in a brown dwarf.  
In this section, we will investigate if the orbital parameters are consistent with Roche lobe overflow.

\subsubsection{Brown dwarf mass}
\label{sec:masses}
The absence of any excess in the optical and the stringent constraints on the companion's luminosity provided by the WIRC $J$-band photometry show that the donors in the two systems are cold brown dwarfs (see Section\,\ref{sec:donors}). This provides a limit on the mass of $M_{\rm donor} <0.08$\,M$_{\odot}$ (assuming solar metallicity; for lower metallicities, the limit is slightly lower). We now explore the constraints on the mass of the donors from the orbital parameters, assuming that the donors are filling their Roche Lobes (we will analyze this assumption further in the next section). 

For each system, we interpret the modulation period in the H$\alpha$ radial velocities as the orbital period of the system. The maximum radial velocity is the maximum velocity along the line of sight. The true maximum velocity depends on the inclination of the system. Following the methodology of \citep{breedt2012}, the corrected maximum velocity is given by
\begin{equation}
K_2=\frac{K_{\rm em}}{1-(1+q)fR_2/a}
\end{equation}
where $K_2$ is the corrected radial velocity of the secondary, $K_{\rm em}$ is the measured radial velocity of the emission line, $q=M_2/M_1$ is the mass ratio of the secondary ($M_2$) and primary ($M_1$), $a$ is the orbital separation, and $fR_2$ is the distance of the centre of light from the centre of mass of the secondary. For this final term, $0\leq f\leq 1$ and $f=0$ represents the case where the emission is uniform across the surface of the secondary. We adopt $f=4/(3\pi)$, as shown by \citet{wade1988} to be appropriate for uniform emission from the day side, i.e., the irradiated side facing the white dwarf. The orbital separation, $a$, can be expressed as
\begin{equation}
    a=\left( \frac{P_{\rm orb}^2G(M_1+M_2)}{4\pi^2} \right)^{1/3}\, ,
\end{equation}
where $P_{\rm orb}$ is the orbital period and $G$ is the gravitational constant. If the secondary is assumed to be undergoing Roche lobe overflow (RLOF), the radius of the secondary, $R_2$, can be assumed to be equivalent to the Roche radius. From \citet{eggleton1983}, this radius is given by the following expression 
\begin{equation}
R_2=a\frac{0.49q^{2/3}}{0.6q^{2/3}+\ln(1+q^{1/3})}.
\end{equation}
Combining the above three equations together, the corrected velocity can be expressed as
\begin{equation}
K_2 = \frac{K_{\rm em}}{1-(1+q)f\frac{0.49q^{2/3}}{0.6q^{2/3}+\ln(1+q^{1/3})}}\, .
\label{eq:K2}
\end{equation}
For WD\,J1820 and WD\,J1907, the corrected velocity, $K_2$, is larger than the measured velocity, $K_{\rm em}$, by $\approx$15\% and $\approx$10\%, respectively.
Rearranging the above equation allows the secondary mass to be expressed as 
\begin{equation}
     M_2 = \left[\frac{4\pi^2P_{\rm orb}}{G} \left(\frac{K_2}{M_12\pi\sin i}\right)^{3}\right]^{-1/2} -M_1\,.
     \label{eq:M2}
\end{equation}
The orbital period, $P_{\rm orb}$, and white dwarf mass, $M_{1}$ have been measured independently by means of spectroscopy and photometry, respectively. The true orbital velocity of the secondary, $K_2$, is given by Eq.\,\eqref{eq:K2}, and depends only on the measured amplitude of the radial velocity, the mass of the primary and secondary, and the covering fraction of H-alpha emission. Since the secondary mass appears on both sides of Eq.\,\eqref{eq:M2} (inside the mass ratio $q$), we solve the equation using a numerical root finding algorithm.

\begin{figure}
	\centering
	\subfloat{\includegraphics[width=0.45\textwidth]{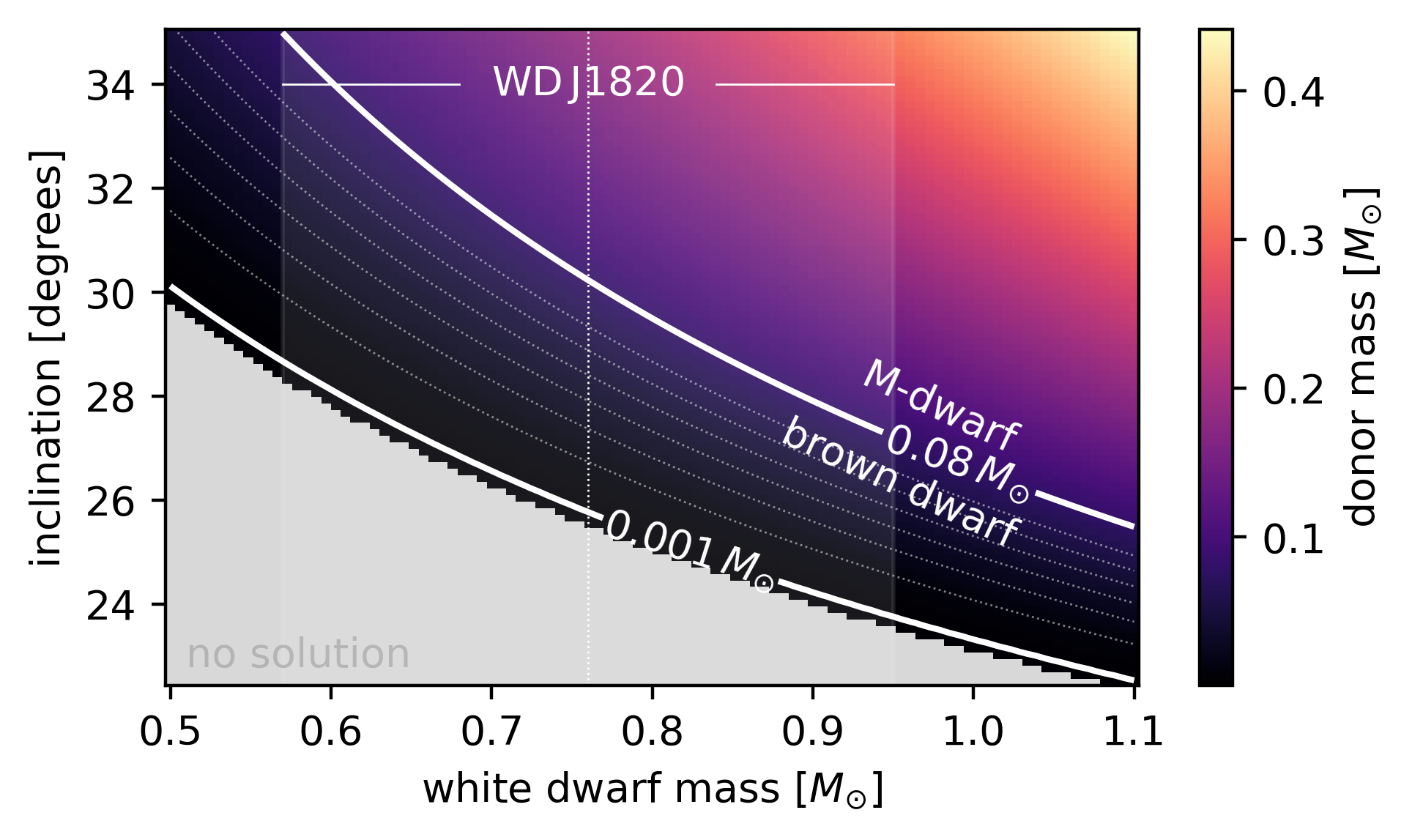}} \\
	\subfloat{\includegraphics[width=0.45\textwidth]{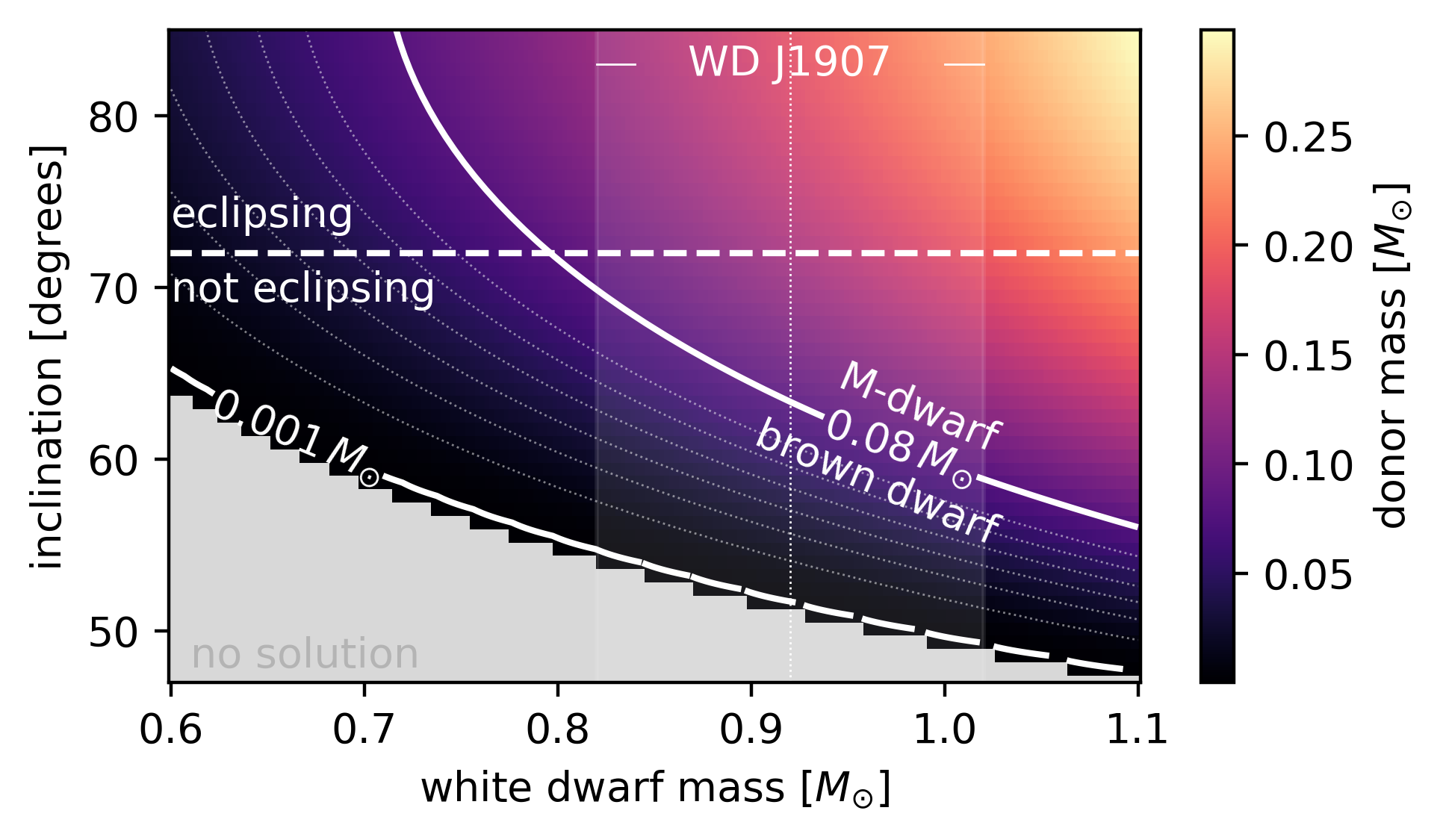}}
	\caption{Dynamical donor mass for the measured orbital period of the two systems studied in this work. The colours indicate the dynamical mass resulting from the solution to Eq.\,\eqref{eq:M2}. We solve this equation for a range of white dwarf masses on the $x$-axis, and a range of inclinations on the $y$-axis. For each white dwarf mass, the inclination has a lower limit, below which no solution can be found (grey shaded region). This limit happens to correspond approximately to a Jupiter mass for the brown dwarf companion. The thick solid white line indicates the upper limit on the mass of a brown dwarf. Below this mass, we show faint white contours at integer multiples of 0.01\,M$_{\odot}$. The vertical white line indicates the white dwarf mass, as inferred from the best-fit parameters of the model atmosphere fit to the SED. The shaded region either side of the vertical line shows the 1$\sigma$ uncertainty on the white dwarf mass measurement. The upper panel pertains to WD\,J1820, whilst the lower panel shows the solutions for WD\,J1907. In the case of WD\,J1907, the range of allowed inclinations implies a larger binary inclination (more edge-on). The horizontal white dashed line indicates an inclination of 72$^{\circ}$, the approximate inclination above which the system is expected to be eclipsing.}
	\label{fg:inclination}
\end{figure}

In Fig.\,\ref{fg:inclination} we show the solution to Eq.\,\eqref{eq:M2}. The colours correspond to the donor mass for a range of white dwarf masses ($x$-axis) and inclinations ($y$-axis). In the grey shaded region no solution can be found; this limit happens to correspond approximately to
a Jupiter mass. The thick solid white line shows the upper limit for the mass of a brown dwarf at 0.08\,M$_{\odot}$. For reference, the thinner solid line indicates a Jupiter mass at 0.001\,M$_{\odot}$. 
We also show the best-fit white dwarf mass and uncertainty for each system, determined from fitting the SED with pure-hydrogen model atmospheres (Section\,\ref{sec:WDfitting}, Table\,\ref{tab:Phot_fitting_optical}). We estimate the dynamical brown dwarf mass by taking the mean mass of all solutions less than 0.08\,M$_{\odot}$, and a white dwarf mass within 1$\sigma$ of the measured value. This yields a dynamical brown dwarf mass of $M_{\rm bd} = 0.04\pm0.03$\,M$_{\odot}$ for WD\,J1820 and $M_{\rm bd} = 0.05\pm0.02$\,M$_{\odot}$ for WD\,J1907. Our approach for deriving the uncertainty on the dynamical mass means that this estimation accounts for the uncertainty on the measured white dwarf mass and the range of allowable inclinations. Although it is not shown in this figure, we remind the reader that the observational lower limit on the inclination (see Section\,\ref{sec:analysis-orbital-inclination}) was derived to be $i>(14\pm2)^{\circ}$ and $i>(28\pm10)^{\circ}$, for WD\,J1820 and WD\,J1907, respectively, also assuming that the emission was coming from the day side of the brown dwarf. The dynamical solutions are fully consistent with these limits.

In the case of WD\,J1907, the range of allowed inclinations implies a larger binary inclination (closer to edge-on). The horizontal white dashed line indicates an inclination of $i$\,=\,72$^{\circ}$, above which the system would be expected to be eclipsing. From optical lightcurves in the ZTF archive, we find no evidence that either system is eclipsing, although these systems are at the edge of the sensitivity of ZTF. Nonetheless, from dynamical arguments, the system would need to have a white dwarf mass less than $M_{\rm wd}<0.8$ to be eclipsing. This mass range is ruled out at 1$\sigma$ by the photometric white dwarf mass, which for WD\,J1907 is $M_{\rm wd}=0.92\pm0.1$\,M$_{\odot}$. 

\subsubsection{Roche lobe overflow}
In this section we consider whether the dynamical masses and radii derived are consistent with the mass-radius relation of non-accreting brown dwarfs. We preface this discussion with the caveat that, as the donor in this system is undergoing mass loss, it is likely that this evolved donor is not in thermal equilibrium, and could thus be significantly inflated.

\begin{figure}
	\centering
	\subfloat{\includegraphics[width=0.46\textwidth]{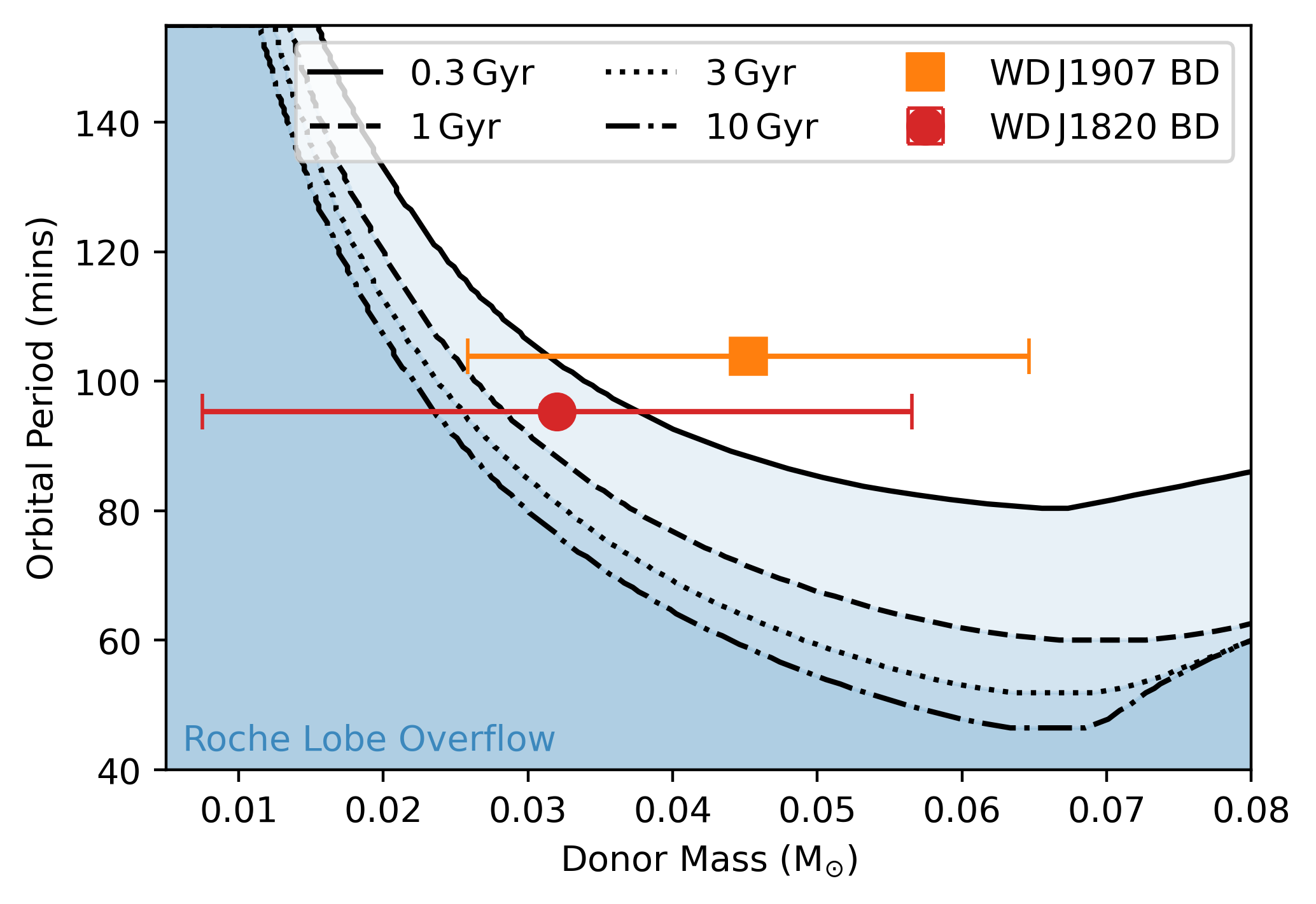}}
	\caption{The orbital period required to achieve Roche lobe overflow, as a function of donor mass. Here it is assumed that the donor is a brown dwarf, and that its mass follows the brown dwarf mass-radius relation. In black, we adopt the M-R relation of \citet{marley2021-Sonora} for four brown dwarf ages; 0.3, 1, 3, and 10\,Gyr. The shaded region beneath each curve indicates the orbital periods for which the secondary Roche lobe would be smaller than an isolate brown dwarf radius, at which point Roche lobe overflow is expected to take place. The data points show the measured orbital period and inferred dynamical brown dwarf mass for WD\,J1820 (red circle) and WD\,J1907 (orange square). We find that both systems have a dynamical mass and radius consistent with a brown dwarf donor filling its Roche lobe. The error bars in dynamical mass are those that were derived in Section\,\ref{sec:analysis-orbital}. The error bars on the period are too small to be seen.}
	\label{fg:RLOF}
\end{figure}

In Fig.\,\ref{fg:RLOF}, we show the orbital period required for a donor to fill its Roche lobe. The black curves show the maximum orbital period for which an isolated brown dwarf of a given mass has a radius larger than the implied Roche radius for a secondary with the dynamical mass and orbital period measured for the two systems. The shaded area under the curves indicates the parameter space consistent with Roche lobe overflow. The dynamical masses that we inferred in the previous section have large uncertainties; however, they are consistent with RLOF being the dominant physical mechanism driving the accretion.

\subsection{Cyclotron emission}
\label{sec:cyclotron}

\begin{figure}
	\centering
	\subfloat{\includegraphics[width=0.88\columnwidth]{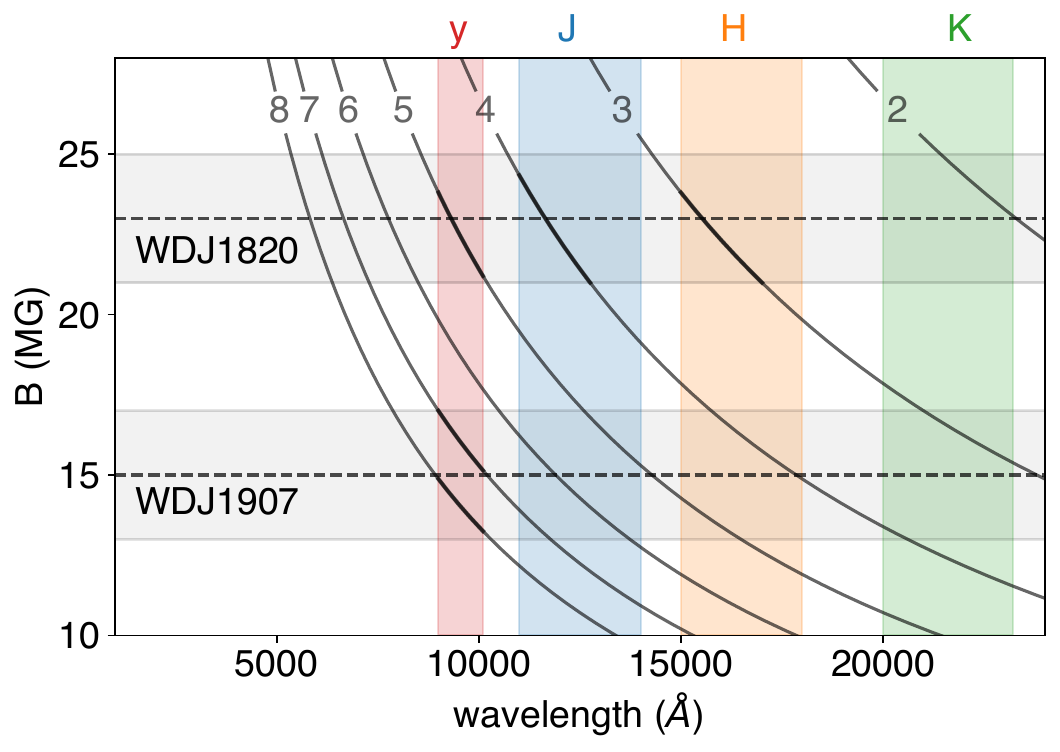}}  
	\caption{Central wavelengths of the first eight cyclotron harmonics for magnetic field strengths within the range measured for the two system presented in this work. If we take the average magnetic field strength of WD\,J1907 (15\,MG), we find that the observed $y$-excess can be explained as the 7$^{\rm th}$ or 8$^{\rm th}$ cyclotron harmonics (the magnetic field at the pole can be a bit stronger, up to twice as much but likely less than that). For WD\,J1820, which has a measured magnetic field strength of 23\,MG, the observed excesses in $y$, J, and H can be explained by the 5$^{\rm th}$, 4$^{\rm th}$, and 3$^{\rm rd}$, harmonics respectively.}
	\label{fg:cyclotron}
\end{figure}

In the case of WD\,J1820, the two epochs of $J$-band photometry show 1.5 magnitudes of variability. In Section\,\ref{sec:donors}, we use the dimmer epoch in $J$-band to constrain the temperature of the brown dwarf donor to be less than about 1100~K (see also Fig.\,\ref{fg:phot_fit_WD+BD} and Table\,\ref{tab:Phot_fitting_WD+BD}). The brighter epoch from UKIDSS shows a strong excess in all three NIR bands $J$, $H$, and $K_{\rm s}$ due to cyclotron emission. 
In Fig\,\ref{fg:cyclotron}, it can be seen that this excess is consistent with emission at the 2$^{\rm nd}$, 3$^{\rm rd}$, and 4$^{\rm th}$ cyclotron harmonics for WD\,J1820, which has an inferred magnetic field of 24\,MG.

To estimate the accretion luminosity emitted through cyclotron cooling, we extract the excess flux in the $J$, $H$ and $K_{\rm s}$ bands. For WD\,J1820, we remove the best-fit WD+BD model from the measured $JHK_{\rm s}$ photometry in Fig.\,\ref{fg:phot_fit_WD+BD}. 
We find fluxes in the $J$, $H$ and $K_{\rm s}$ bands of 2.4$\times10^{-14}$, 2.2$\times10^{-14}$, and 3.3$\times10^{-15}$ erg/s/cm$^2$, respectively. The summed cyclotron flux is thus $5\times10^{-14}$\,erg/s/cm$^2$. This is over an order of magnitude larger than the integrated flux measured in the 0.25--10.0\,keV band with XMM-Newton (see Table\,\ref{tab:bxa-bb-apec-Mdot}). However, we note that both the X-ray and NIR integrated fluxes are lower limits, and more observations would be be needed to characterise the total accretion-induced luminosity at EUV and optical/NIR wavelengths.

For WD\,J1907, we identify an excess in PS1-$y$, which we also interpret as evidence of cyclotron emission. In this case, we only have one $J$-band epoch, and no available photometry at longer wavelengths. Nonetheless, the $J$-band epoch provides a stringent constraint on the temperature and radius of donor, suggesting that the donor must be a brown dwarf cooler than 1000\,K, consistent with a mid-late T-dwarf of at least type T6 \citep{faherty2016}. This reveals an excess in PS1-$y$ of at least 0.5 magnitudes. In Fig.\,\ref{fg:cyclotron}, we show the central wavelengths of the cyclotron harmonics for a range of magnetic field strengths. At the measured magnetic field strength of 15\,MG, the $y$-excess would be consistent with emission from the 7$^{\rm th}$ or 8$^{\rm th}$ cyclotron harmonics. Given that we only have evidence of a significant excess in one passband, we don't provide an estimate of the cyclotron flux, but future time-series photometric or spectroscopic observations at optical and NIR wavelengths would allow to robustly constrain the cyclotron flux.

\subsection{Evolutionary status}
\label{sec:evol}
\begin{figure}
	\centering
 	\subfloat{\includegraphics[width=0.48\columnwidth]{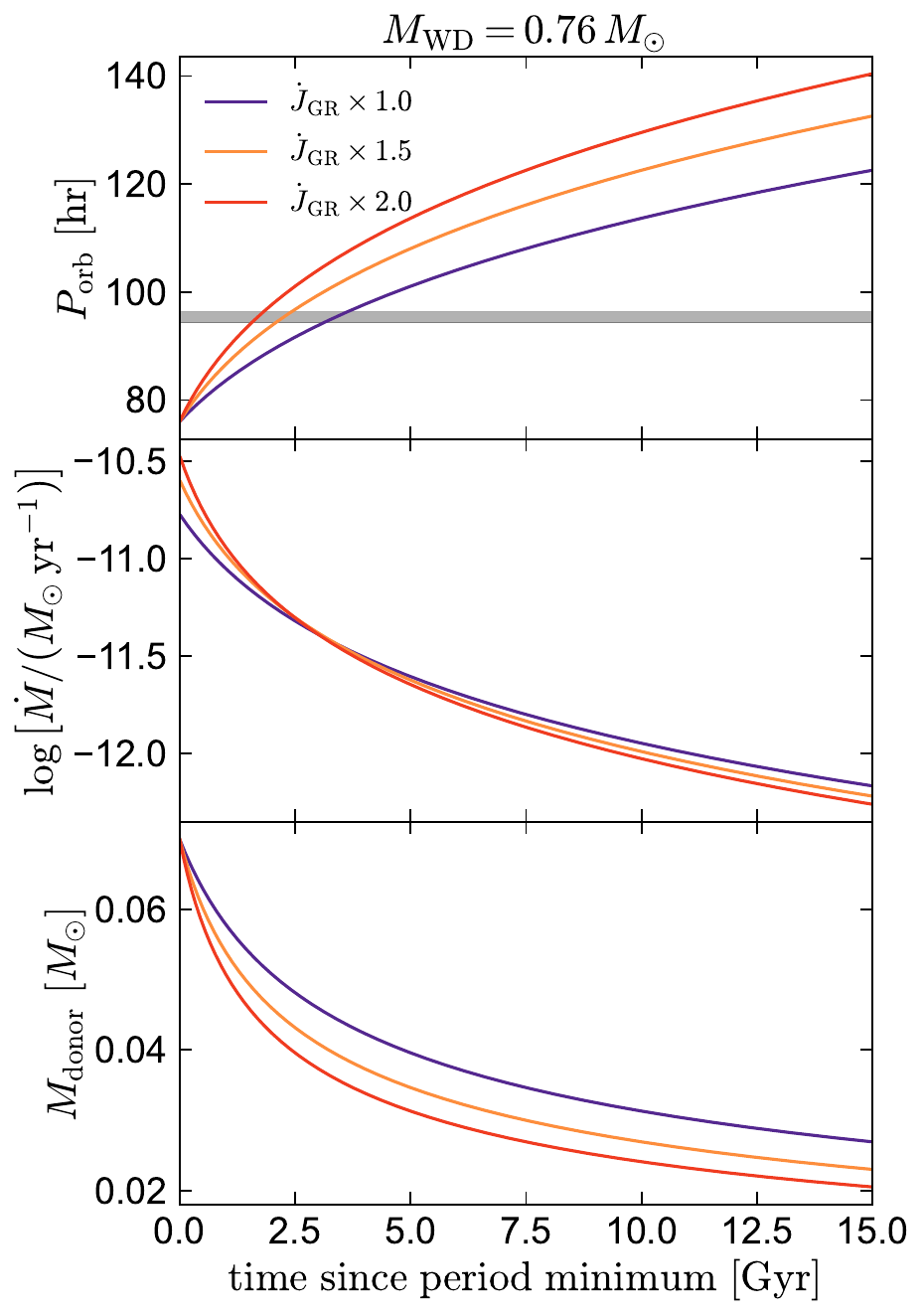}}
	\subfloat{\includegraphics[width=0.48\columnwidth]{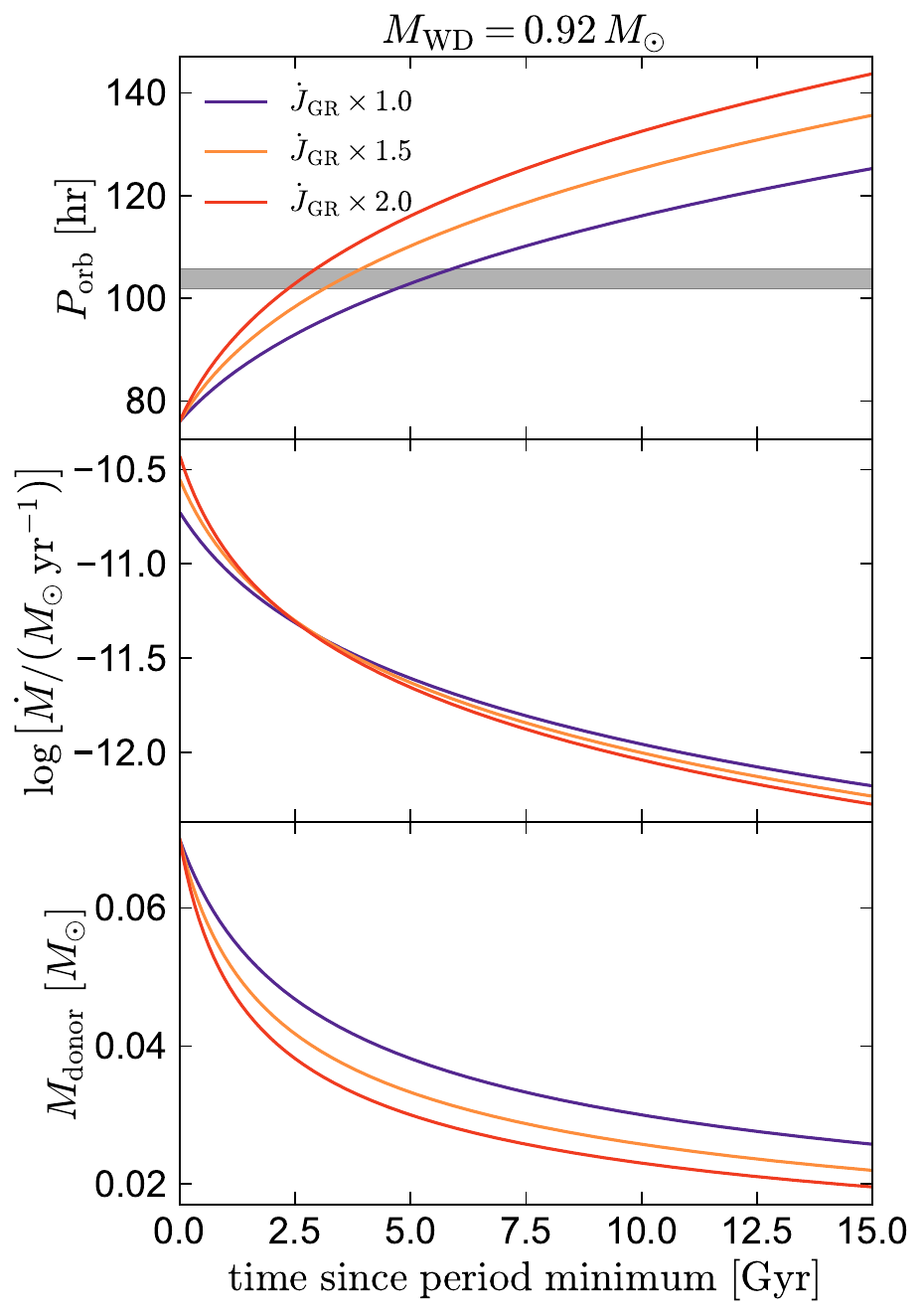}}
  
	\caption{Predicted evolution of the period, accretion rate and donor mass for the two systems, WD\,J1820 (left) and WD\,J1907 (right), as a function of time after the period bounce. The calculation assumes that the angular momentum loss is driven by gravitational wave emission only ($\dot{J}_{\rm{GR}}$, dark purple lines), or that there is an enhanced angular momentum loss, parametrized as $1.5\times\dot{J}_{\rm{GR}}$ (ochra lines) and as $2\times\dot{J}_{\rm{GR}}$ (red lines). The shaded grey region indicates the 1$\sigma$ confidence measurements of the two periods.}
	\label{fg:post-bounce-evolution}
\end{figure}

Angular momentum losses after the period bounce are often assumed to be driven by gravitational wave emission only. As the binary evolves after the period minimum, the angular momentum losses are driving the ongoing mass transfer, keeping the brown dwarf radius close to the Roche radius. Given an assumption on the value of the angular momentum change in the binary, $\dot{J}(t)$, the evolution of the period and average accretion rate with time can be calculated, as well as the evolution of the donor mass \citep{deloye2007,knigge2011}. In Figure\,\ref{fg:post-bounce-evolution}, we plot the evolution assuming that $\dot{J}$ is governed only by gravitational waves emission ($\dot{J}_{\rm{GR}}$); additionally, to show the effect of enhanced losses by other mechanisms (like for example residual magnetic braking), we also show the evolution assuming $\dot{J}$ to be larger by 1.5 and 2 times with respect to $\dot{J}_{\rm{GR}}$. The shaded bands in the figure show the 1\,$\sigma$ confidence measurements on the periods of the two bouncers as measured via the H$\alpha$ emission. Assuming only gravitational waves losses, we can see that to reach the current periods, the two systems would have evolved past the period minimum at least 3 Gyr ago for WD\,J1820, and about 5 Gyr ago for WD\,J1907. The predicted donor masses at this evolutionary stage are about 0.04\,M$_\odot$ for both systems, consistent with the dynamical masses we derived in Section\,\ref{sec:masses}. 

At this point in the evolution, the expected accretion rate is approximately $\approx$\,$10^{-11.5}$\,M$_{\odot}$/yr for both systems. The temperature of the white dwarf accretor has been used before as a proxy for the secular accretion rate in CVs \citep{townsley2002,townsley2003,townsley2004,townsley2005,townsley2009}.  
Equation 2 of \cite{townsley2009} provides a predicted accretion rate for a given white dwarf temperature for a typical CV. Given the effective temperatures of our systems, the predicted accretion rate is on the order $10^{-11}$\,M$_{\odot}$\,yr$^{-1}$. This is broadly consistent with the accretion rate predicted by the evolutionary models in Fig.\ref{fg:post-bounce-evolution}, given the measured periods. 

\begin{figure}
	\centering
	\subfloat{\includegraphics[width=0.9\columnwidth]{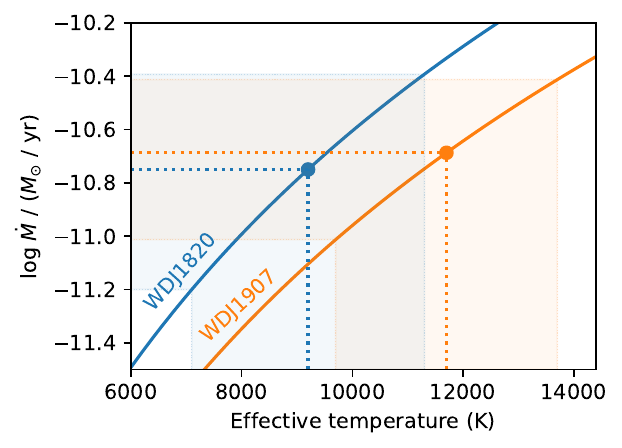}}  
	\caption{Accretion rate as a function of white dwarf effective temperature as predicted by Equation 2 of \citet{townsley2009}. The predicted accretion rate for both systems presented in this work are around $10^{-11}$\,$M_{\odot}$\,yr$^{-1}$, 3 dex higher than that measured in the \textit{XMM-Newton} observations. However, as discussed in Section\,\ref{sec:cyclotron}, cyclotron emission in the near-infrared accounts for at least an order of magnitude more accretion-induced luminosity; additionally, we might have caught these systems in a low state, in which their accretion rate is much lower than average.}
	\label{fg:Mdot-Eq2-townsley-2009}
\end{figure}

The predicted accretion rates are around 2.5 dex higher than measured via our X-ray analysis (see Table\,\ref{tab:bxa-bb-apec-Mdot}). These measured low rates could indicate that the accretion is instead driven by weak winds from the brown dwarfs \citep{schreiber2023}, and not by Roche lobe overflow. However, as we discussed in Section\,\ref{sec:accrate}, the measured X-ray accretion rate is formally a lower limit, as the accretion-induced luminosity could be emitted at other wavelengths and through cyclotron emission. 
We do have strong evidence of cyclotron emission (see Section\,\ref{sec:cyclotron}), consistent with the first few cyclotron harmonics. This suggests that some of the accretion power is being radiated in the NIR through cyclotron cooling, especially for WD\,J1820, for which the cyclotron emission accounts for at least an order of magnitude more accretion-powered flux compared to the X-ray. Time-series NIR spectroscopy and photometry would be required to better constrain the NIR accretion-induced luminosity. Also, as the case of EF Eri shows, we might have caught these systems in a low state, and their average accretion rate might be higher, as hinted by their surface temperatures. 

\subsection{Currently known candidates}
\label{sec:pop}

\begin{table}
	\centering
	\caption{ \label{tab:pbs} \small Physical parameters of the known magnetic period bouncers candidates and of the two newly-discovered long-period polars presented in this work. We include the following parameters: distance ($d$), mean surface magnetic field strength ($B_{\rm WD}$), orbital period ($P_{\rm orb}$), X-ray luminosity ($L_{\rm X}$), and spectral type of the donor (SpT$_{\rm don}$). The right-pointing arrows in the donor spectral type column indicate an upper limit on the spectral type, i.e., consistent with the quoted spectral type or later.
 }
 \vspace{5pt}
    \footnotesize
        \setlength\tabcolsep{3.0pt} 
 	\begin{tabular}{lcccclc} 
		\hline\hline
          & d & $B_{\rm WD}$ & $P_{\rm orb}$ & $L_{\rm X}$ & SpT$_{\rm don}$ & Ref. \\
          & [pc] & [MG] & [min] & [$10^{29}$erg/s] &  \\
		\hline
            EF Eri	    &160 	& 20  &	81	& 2	& M9$\rightarrow$ & (1,2)   \\
            SDSS\,1250	& 132	& 20  & 87	& 1.6$\pm$0.3	& M8 & (3,4)	  \\ 
            V379 Vir	&155	& 7   &	88	& 3.2$\pm$0.5	& L5--L8 & (3,4)  \\ \vspace{3pt}
            SDSS\,1514	&181	& 36  &	89	& 3.2$^{+3.6}_{-2.7}$	& L3 & (3,4)    \\ \vspace{3pt}
            SMSS\,J1606	&108	& 30  &	92	& 	& L8$\rightarrow$ & (5)    \\ \vspace{3pt}
            1eRASS\,J05472	& 132	& $>$\,10  &	94	& 1.8$\pm1.0$	& M6$\rightarrow$ & (6)    \\             
            WD\,J1820	&213	& 24  &	95	& 0.3$\pm$0.1	&T5$\rightarrow$ & this work\\ \vspace{3pt}
            WD\,J1907  	&243	& 15  & 104	& 1.9$\pm$0.3	& T6$\rightarrow$ & this work \\	

            ZTF\,0146 	&56	    & 89  & 123	&  & T9$\rightarrow$ & (7)   \\

		\hline\hline
	\end{tabular}
\footnotesize
References: (1) \citet{ferrario1996}, (2) \citet{schwope2007}, (3) \citet{breedt2012}, (4) \citet{munoz-giraldo2023}, (5) \citet{kawka2021}, (6) \citet{rodriguez2024}, (7) \citet{guidry2021}
\end{table}

We list all the known candidate and confirmed magnetic period bouncers from the literature and this work in Table\,\ref{tab:pbs}. All the objects in the table are characterized by ongoing accretion, confirmed through the detection of X-ray emission or of cyclotron emission, and very late-type companions. The low-mass donors of these systems have either no or very weak coronal X-ray emission, and therefore the detected X-rays are enough to confirm accretion.

The most well known system is EF Eridani (EF Eri). It was first detected in X-rays in the 1980s with the \textit{HEAO 1} and \textit{Einstein} observatories \citep{white1981}, and it looked like a prototypical polar. In 1997, however, EF Eri fell into a low state \citep{wheatley98}, from which it emerged in late 2022. XMM-Newton observations in its low state \citep{schwope2007} measured an X-ray luminosity $L_{\rm X}=2\times10^{29}$\,erg/s, about 3 dex lower than in the high state. The plasma temperature was also found to be significantly lower at 2.0\,keV, compared to 10\,keV in the high state. The white dwarf accretor has a temperature of about 10,000\,K, but the system shows a variable excess in the UV, which has been modeled as a 20,000\,K hot spot \citep{szkody2006,schwope2007}. 
The period of EF Eri is very close to the period minimum \citep[81 minutes;][]{beuermann2000-EF-Eri}, but its nature as period bouncer is still unclear, as the donor has not been confirmed to be sub-stellar. EF Eri shows IR variability due to cyclotron emission and the minimum in the $J$-band has been used to constrain the contribution from the companion. Based on this constraint, \citet{schwope2010} suggested that the donor is a sub-stellar object; however, the revised distance obtained with Gaia DR3 increased the estimated absolute magnitude of the donor to above the sub-stellar limit \citep{munoz-giraldo2023}. 

SDSS\,J125044.42+154957.4 (SDSS 1250) was found in a cross-match between SDSS and the United Kingdom Infrared Telescope (UKIRT) Infrared Deep Sky Survey (UKIDSS) to have a late M-dwarf companion \citep{breedt2012}, which shows variable H$\alpha$ on the orbital period due to irradiation. Accretion in this system was confirmed by the detection of X-rays from the all-sky survey eROSITA \citep{predehl2021,munoz-giraldo2023}. Again, the low-accretion rate and late-type of the donor indicate that this system is either moving toward the period minimum or is a period bouncer.

SMSS\,J1606$-$1000 was discovered among a selection of blue objects from the SkyMapper survey \citep{kawka2021}; phase-resolved optical spectroscopy reveals a 30~MG magnetic white dwarf with H$\alpha$ and H$\beta$ emission varying in radial velocity with an orbital period of 92 minutes. An SED analysis of near-IR data constrains the companion to be later than a L8 brown dwarf, and shows the presence of cyclotron emission. The system looks quite similar to the two objects analysed in this paper, although no X-ray observations are available.

SDSS\,J151415.65+074446.5 (SDSS\,1514) was also discovered in a cross-match between SDSS and UKIDSS \citep{breedt2012}, while SDSS\,J121209.31+013627.7 (V379 Vir) was detected as an X-ray source with Swift \citep{schmidt2005a,burleigh2006}. They have very similar periods, slightly longer than the period minimum, and in both systems, IR photometry strongly constrains the contribution of the donor, indicating the presence of a brown dwarf companion \citep{farihi2008,breedt2012}. The sub-stellar nature of the donors, contrary to the previous two system, exclude the possibility that these two systems may still be moving toward the period minimum, and characterises them as period bouncers. They both were later targeted with XMM-Newton \citep{stelzer2017,munoz-giraldo2023}, showing plasma temperatures of $kT\approx3$\,keV and X-ray luminosities in of $L_{\rm X}$\,$\sim$\,$10^{29}$\,erg/s, similar to the new systems presented in this work.

1eRASS\,J054726.9+132649 (1eRASS\,J0547) was recently discovered by \citet{rodriguez2024} in a cross-match between the eRASS1 source catalogue \citep{merloni2024} and Gaia. This system has an orbital period of 94\,min, as measured via a Lomb-Scargle periodogram analysis of the ZTF $g$- and $r$-band light curves. The authors do not provide a measurement of the magnetic field or donor temperature. However, the presence of resolved Zeeman splitting in the optical spectrum implies a magnetic field strength of $B$\,$\gtrapprox$\,1\,MG, but the absence of a disk implies $B$\,$\gtrapprox$\,10\,MG. For the donor, the authors rule out effective temperatures in excess of $\approx$\,2,900\,K, or spectral types earlier than M6.

To this list, we add ZTF\,J014635.73+491443.0 (ZTF\,J0146), which has not yet been targeted by X-ray observations, but which shows ongoing accretion thanks to strong cyclotron emission in the optical. The candidate polar was identified by \citet{guidry2021} in the Zwicky Transient Facility archive thanks to its large optical modulation due to varying cyclotron emission on its 123-minute orbital period.  WISE infrared photometry places stringent constraints on the donor: W1 and W2 are consistent with 
the white dwarf spectrum, allowing only a $\sim$500\,K or cooler brown dwarf to hide at these wavelengths \citep{hakala2022}. The object has fallen into three slew observations with \textit{XMM-Newton}, which provide an upper limit on the X-ray flux of $F_{\rm X}<2.2\times10^{-12}$\,erg\,s$^{-1}$\,cm$^{-2}$. At a distance of only 56\,pc, the limiting luminosity is then $L_{\rm X}<8\times10^{29}$\,erg/s. This is consistent with the observations of the two new polars presented in this work, which have 100-min periods and X-ray luminosities less than $10^{29}$\,erg/s. We thus expect that a reasonably deep pointing of $\approx$50\,ks with XMM-Newton should be sufficient to detect and characterise X-ray emission at this system in order to confirm it as a low accretion rate polar.

We would also like to mention that recently, \citet{galiullin2024} discovered an eclipsing period bouncer in a cross-match between the eRosita and the ZTF catalog, SRGeJ041130.3+685350. The absence of a disk in this system, together with a soft X-ray spectrum, hint to the white dwarf accretor being magnetic; however, we do not include it in our list as the evidence is not conclusive for the primary to be a magnetic white dwarf.

\subsection{Space density}
The empirical space density of cataclysmic variables is a vital diagnostic in characterising the population and evolution of CVs. Evolutionary models have predicted larger space densities of CVs than is currently observed \citep{breedt2012,pala2017,inight2023}. In particular, it has been predicted that 40--70\% of CVs should have evolved through the period bounce \citep{kolb1993,goliasch2015,belloni2020}, but this large predicted population has been mostly elusive so far. Many observational studies have characterized the space density of CVs, including the different sub-types \citep{pala2017,inight2023,rodriguez2024}. \citet{inight2023} recently provided an empirical determination of the space density of CVs, finding that only 3.4 per cent of known CVs have undergone a period bounce; in contrast, \citet{rodriguez2024} found a higher value of 25 per cent. Even this higher estimate is inconsistent with predictions from CV evolutionary models. In the following, we consider the implications for space density from the discoveries presented in this work.

The Gaia white dwarf catalogue has fairly uniform sky coverage within $\approx$200\,pc. Both of the systems studied in this work have distances around 250\,pc. 
The 4XMM-DR13 source catalogue includes serendipitously detected sources across the 1,328 sq. deg of sky that has been subjected to targeted XMM-Newton observations. This sky coverage corresponds to 3.2\% of the total sky area. We can thus make a crude estimate that the total sky should contain $2/0.032=62\pm44$ period-bounce polars with comparable X-ray luminosity to the ones presented here, where the uncertainty is estimated from Poisson statistics for the sample size of $2\pm\sqrt2$. We use this simple estimate and associated uncertainty to calculate the space density of period-bounce polars. Both polars presented in this study are within 250\,pc, so we adopt this as the limiting distance. In characterising the space density of CVs, \citet{pretorius2007} adopted three scale heights for different evolutionary stages: 120, 260, and 450\,pc for long-period systems ($P_{\rm orb}$\,$>$\,$3$\,h), ``normal'' short-period systems ($P_{\rm orb}$\,$<$\,$3$\,h) and period bouncers, respectively. We adopt the latter scale height, both because we are interested in calculating the space density of period bounce magnetic systems, and because it is the most conservative of the three. For the scale height, $h$, of the volume distribution of a sample in the Galactic disk, the density of systems as a function of the height, $z$, off the Galactic plane is described by 
\[
\rho(z) = \rho_0 \, e^{-|z|/h}\,,
\]
where $\rho_0$ is the space density. The space density is estimated by dividing the sample size by the limiting volume. The effective volume, given the scale height is given by the expression 
\[
V_{\text{eff}} = \int\limits_0^{2\pi} \int\limits_0^{\pi} \int\limits_0^R \rho(r \sin\theta) \, r^2 \sin\theta \, dr \, d\theta \, d\phi\,,
\]
where $R$ is the limiting distance.
For a scale height of 450\,pc and an expected sample size of $62\pm44$, we estimate the space density of period-bounce magnetic CVs to be $\rho_0$\,$=$\,$(1.3\pm0.9)\times10^{-6}$\,pc$^{-3}$. 

Space density estimates for the entire population of CVs (both magnetic and non-magnetic) have been found to be $\rho_0$\,$=$\,$4.8\times10^{-6}$\,pc$^{-3}$ \citep{pala2020} and $\rho_0$\,$=$\,$5.8\times10^{-6}$\,pc$^{-3}$ \citep{inight2023}, for assumed Galactic scale heights of 260 and 280\,pc, respectively, where the assumed Galactic scale height represents a major uncertainty in the estimates of space densities. In the case of magnetic CVs, \citet{pretorious2013} estimated the space density based on the X-ray flux-limited ROSAT Bright Survey \citep[RBS;][]{schwope2000-RBS}. The authors found a combined space density of magnetic CVs, including polars and intermediate polars to be $\rho_0$\,$=$\,$8^{+4}_{-2} \times 10^{-7}$pc$^{-3}$. \citet{rodriguez2024} also provide an estimate of the space density of all magnetic CVs to be $\rho_0$\,$=$\,$1.3\pm0.5 \times10^{-6}$\,pc$^{-3}$. This is comparable to the estimate of space density we provide in this work for the 2 period bounce magnetic CVs detected serendipitously with XMM alone, suggesting that the evolved magnetic CVs may account for $(50\pm40)$\% of the total population of magnetic CVs. The low numbers implies a large uncertainty in our estimate; however, this is consistent with the predicted 40--70\% from evolutionary models.

We now briefly consider whether our estimated space density of period-bounce magnetic CVs is consistent with the detections or lack of thereof of such systems in all-sky X-ray surveys. The all-sky ROSAT X-ray survey had a flux limit of $F_{\rm X}$\,$\gtrapprox$\,$3\times10^{-13}$\,erg\,s$^{-1}$\,cm$^{-2}$ in the 0.12--2.48 keV band \citep{boller2016}, whilst the systems studied in this work have X-ray fluxes of $F_{\rm X}$\,$\sim$\,$10^{-15}$\,erg\,s$^{-1}$\,cm$^{-2}$ and $F_{\rm X}$\,$\sim$\,$10^{-14}$\,erg\,s$^{-1}$\,cm$^{-2}$, for WD\,J1820 and WD\,J1907, respectively (see Table\,\ref{tab:bxa-bb-apec-Mdot}). In order for both systems to have been detected by ROSAT, they would need to be located an order of magnitude closer, i.e. at $\approx$\,10--40\,pc. From our space density estimate, the volume defined within 25\,pc of the Sun should contain $\approx$\,$0.3$\,$\pm$\,$0.1$ such systems. Thus, our estimated space density is consistent with the non-detection of such systems in ROSAT. For eROSITA, the limiting flux in the 0.2--2.3\,keV band has been shown to be $F_{\rm X}$\,$\gtrapprox$\,$5\times10^{-14}$\,erg\,s$^{-1}$\,cm$^{-2}$  
\citep{merloni2024}. At this limiting flux, the systems presented in this work should be detectable up to distances of $\approx$\,30--100\,pc, considering the measured X-ray fluxes for WD\,J1820 and WD\,J1907. Within this volume, and accounting for the fact that publicly available eROSITA data covers 50\% of the sky, our calculation predicts a sample of 0--2 period-bounce magnetic CVs detectable with eROSITA. \citet{rodriguez2024} recently reported the detection of 2 magnetic period-bouncers in the eRASS1 catalogue; one known magnetic period bouncer (SDSS\,J1250) and one newly discovered (1eRASS J0547, see Table\,\ref{tab:pbs}). This sample size is consistent with the estimate based on the 2 serendipitous discoveries presented in this work.

\section{Conclusions}
\label{sec:conclusions}
In this work we have presented two new magnetic period-bounce CVs, discovered serendipitously via their X-ray emission reported in the 4XMM-DR13 source catalogue, and their membership of the Gaia white dwarf catalogue. We obtained time-series optical spectroscopy to characterise the nature of the white dwarf candidates, confirming these as highly-magnetized white dwarfs with irradiated companions. The presence of X-ray emission confirms these as accreting white dwarfs, whilst near-infrared photometry demonstrates that the white dwarfs are accreting from cool brown dwarf companions and show variable cyclotron emission. The discovery of these two systems, significantly increases the sample of period-bounce magnetic CVs with confirmed sub-stellar companions. We have provided a summary of the characteristics of all known magnetic period bouncers in section\,\ref{sec:pop}. 

The two new systems studied in this work have periods of 95 and 104 minutes. Such periods are expected to take $\sim$\,3--5\,Gyr to reach after the period minimum, characterizing these as the most evolved period-bounce magnetic CVs with detected X-ray emission. The  two systems were discovered via their X-ray emission, published in the 4XMM-DR13 source catalogue. Given that the source catalogue includes sources detected across just 3\% of the sky, the discovery of two systems in such a small fraction of sky offers the tantalising possibility that the space density of period-bounce CVs may be significantly higher than current empirical determinations have suggested. It has long been predicted that period-bounce CVs should account for 40--70\% of the population, but current studies have placed the estimates at closer to 3--25\% \citep{munoz-giraldo2024-period-bounce,rodriguez2024}. Our newly discovered systems effectively double the inferred space density of magnetic, period-bounce CVs, possibly resolving this long-standing discrepancy, at least for the magnetic sample. This result suggests that a large number of similar systems could be detected with targeted XMM observations, or all-sky X-ray surveys such as eROSITA. We also detect infrared variability, evidence of cyclotron emission, in one of the two systems. This suggests that all-sky infrared time-domain surveys such as the Spectro-photometer
for the History of the Universe, Epoch of Reionization, and Ices Explorer \citep[SPHEREx;][]{spherex}, and even the \textit{Nancy Grace
Roman Space Telescope} \citep{roman}, with its much smaller footprint but higher sensitivity, may be effective discovery machines for identifying more of these systems through IR variability to be followed up with optical and X-ray observations.

\section*{Acknowledgements}
We thank Matthias Schreiber for his insightful comments. Support for this work was provided by NASA through the NASA Hubble Fellowship grant HST-HF2-51527.001-A awarded by the Space Telescope Science Institute, which is operated by the Association of Universities for Research in Astronomy, Inc., for NASA, under contract NAS5-26555. IC was also supported by NASA through grants from the Space Telescope Science Institute, under NASA contracts NASA.22K1813, NAS5-26555 and NAS5-03127.  
This project has received funding from the European Research Council (ERC) under the European Union’s Horizon 2020 research and innovation programme (Grant agreement No. 101020057). 
This research was supported in part by grant NSF PHY-1748958 to the Kavli Institute for Theoretical Physics (KITP).
PJW acknowledges support from the UK Science and Technology Facilities Council (STFC) through consolidated grants ST/T000406/1 and ST/X001121/1.
RA was supported by NASA through the NASA Hubble Fellowship grant \#HST-HF2-51499.001-A awarded by the Space Telescope
Science Institute, which is operated by the Association of Universities for Research in Astronomy, Incorporated, under NASA contract NAS5-26555.

This research has made use of data obtained from the 4XMM XMM-Newton Serendipitous Source Catalog compiled by the 10 institutes of the XMM-Newton Survey Science Centre selected by ESA.
This work has made use of data from the European Space Agency (ESA) mission
{\it Gaia} (\url{https://www.cosmos.esa.int/gaia}), processed by the {\it Gaia}
Data Processing and Analysis Consortium (DPAC,
\url{https://www.cosmos.esa.int/web/gaia/dpac/consortium}). Funding for the DPAC
has been provided by national institutions, in particular the institutions
participating in the {\it Gaia} Multilateral Agreement.
The Pan-STARRS1 Surveys (PS1) and the PS1 public science archive have been made possible through contributions by the Institute for Astronomy, the University of Hawaii, the Pan-STARRS Project Office, the Max-Planck Society and its participating institutes, the Max Planck Institute for Astronomy, Heidelberg and the Max Planck Institute for Extraterrestrial Physics, Garching, The Johns Hopkins University, Durham University, the University of Edinburgh, the Queen's University Belfast, the Harvard-Smithsonian Center for Astrophysics, the Las Cumbres Observatory Global Telescope Network Incorporated, the National Central University of Taiwan, the Space Telescope Science Institute, the National Aeronautics and Space Administration under Grant No. NNX08AR22G issued through the Planetary Science Division of the NASA Science Mission Directorate, the National Science Foundation Grant No. AST–1238877, the University of Maryland, Eotvos Lorand University (ELTE), the Los Alamos National Laboratory, and the Gordon and Betty Moore Foundation.
This work is based in part on data obtained as part of the UKIRT Infrared Deep Sky Survey.
This research made use of hips2fits,\footnote{https://alasky.cds.unistra.fr/hips-image-services/hips2fits} a service provided by CDS, and of astropy \citep{astropy}.

\section*{Data Availability}
Upon request, T.C. will provide the reduced WIRC and Keck/LRIS data. The data from Gaia, Pan-STARSS, UKIDDS, and \textit{XMM-Newton} are already in the public domain, and they are readily accessible in the Gaia, Pan-STARSS and UKIDDS catalogues, and in the \textit{XMM-Newton} Science Archive.




\bibliographystyle{mnras}
\bibliography{mybib} 

\appendix

\section{X-ray spectral properties}
In Tables\,\ref{tab:app-bxa-1820}\,\&\,\ref{tab:app-bxa-1907} in this section we provide the 68\% and 90\% confidence limits on the X-ray properties of the two systems studied in this work. We include the integrated flux over two energy ranges, namely a soft band which we define as 0.25--2.0\,keV and a broad band defined as 0.25--10.0\,keV. The two other derived X-ray properties -- luminosity and accretion rate -- are derived following the methodology presented in Section\,\ref{sec:results-xray-fits}.

\begin{table*}
    \centering

    \caption{\label{tab:confidence-interval-Flux}\label{tab:confidence-interval-luminosity}\label{tab:confidence-interval-Mdot}\textbf{Best-fit X-ray properties of WD\,J1820}. The best-fit fluxes are computed across two spectral energy ranges using the one-temperature (\texttt{apec}) model with solar abundances\citep{asplund2009}. 
    The 68\% and 90\% confidence intervals on the flux are derived by sampling the parameter posteriors from the X-ray fitting with BXA \citep{buchner2016-bxa}.
    From the fluxes ($F_{\rm X}$), the X-ray luminosity ($L_{\rm X}$) was derived using $L_{\rm X}=4\pi d^{2} F_{\rm X}$.
    The best-fit X-ray accretion rate ($\dot{M}_{\rm X}$) derived using $\dot{M}_{\rm X}=2L_{\rm X}R_{\rm WD}/ (G M_{\rm WD})$.}
    \label{tab:app-bxa-1820}
    \begin{tabular}{l | l | c c c c c} 
    \hline \hline
     \multirow{2}{*}[0.8ex]{X-ray measurement} & Energy &  \multirow{2}{*}[0.8ex]{Best-fit} & \multicolumn{2}{c}{68\%} & \multicolumn{2}{c}{90\%} \\ 
     &   (keV) &  & low & high & low & high \\ [0.5ex]
    \hline
    
\multirow{2}{*}[-0.2ex]{\shortstack[l]{Flux \\ ($10^{-15}\,\mathrm{erg\,s^{-1}\,cm^{-2}}$)}} 						
	& 0.25--2.0 &	\textbf{1.44} &	0.91 &	2.04 &	0.53 &	2.40 \\
	& 0.25--10.0 &	\textbf{2.39} &	1.29 &	3.97 &	0.77 &	5.08 \\ [1ex]
\hline						
						
   \multirow{2}{*}[-0.2ex]{\shortstack[l]{Luminosity \\ ($10^{28}\,\mathrm{erg\,s^{-1}}$)}} 						
	& 0.25--2.0 &	\textbf{0.74} &	0.47 &	1.05 &	0.27 &	1.23 \\ 
	& 0.25--10.0 &	\textbf{1.23} &	0.66 &	2.03 &	0.40 &	2.61 \\ [1ex]
\hline						
						
   \multirow{2}{*}[-0.2ex]{\shortstack[l]{Accretion rate \\ ($10^{11}\,\mathrm{g\,s^{-1}}$)}}						
	& 0.25--2.0 &	\textbf{1.10} &	0.69 &	1.56 &	0.40 &	1.83 \\ 
	& 0.25--10.0 &	\textbf{1.82} &	0.98 &	3.03 &	0.59 &	3.88 \\ [1ex]
\hline						
						
   \multirow{2}{*}[-0.2ex]{\shortstack[l]{Accretion rate \\ ($10^{-14}\,\mathrm{M_{\odot}\,yr^{-1}}$)}}						
	& 0.25--2.0 &	\textbf{0.17} &	0.11 &	0.25 &	0.06 &	0.29 \\ 
	& 0.25--10.0 &	\textbf{0.29} &	0.16 &	0.48 &	0.09 &	0.62 \\ [1ex]

\hline
\hline    
    \end{tabular}
    \\
    \justifying
    \vspace{15pt}
    \noindent  
\end{table*}

\begin{table*}
    \centering
    \caption{\label{tab:confidence-interval-Flux}\label{tab:confidence-interval-luminosity2}\label{tab:confidence-interval-Mdot2}{Similar to Table\,\ref{tab:confidence-interval-luminosity} but for WD\,J1907}. }
    \label{tab:app-bxa-1907}
    \begin{tabular}{l | l | c c c c c} 
    \hline \hline
     \multirow{2}{*}[0.8ex]{X-ray measurement} & Energy &  \multirow{2}{*}[0.8ex]{Best-fit} & \multicolumn{2}{c}{68\%} & \multicolumn{2}{c}{90\%} \\ 
     &   (keV) &  & low & high & low & high \\ [0.5ex]
    \hline
    
\multirow{2}{*}[-0.2ex]{\shortstack[l]{Flux \\ ($10^{-15}\,\mathrm{erg\,s^{-1}\,cm^{-2}}$)}} 						
	& 0.25--2.0 &	\textbf{4.83} &	4.05 &	5.76 &	3.58 &	6.39 \\
	& 0.25--10.0 &	\textbf{11.19} &	8.36 &	14.62 &	6.82 &	16.86 \\ [1ex]
\hline						
						
   \multirow{2}{*}[-0.2ex]{\shortstack[l]{Luminosity \\ ($10^{28}\,\mathrm{erg\,s^{-1}}$)}} 						
	& 0.25--2.0 &	\textbf{3.17} &	2.66 &	3.78 &	2.35 &	4.20 \\ 
	& 0.25--10.0 &	\textbf{11.37} &	8.50 &	14.86 &	6.93 &	17.12 \\ [1ex]
\hline						
						
   \multirow{2}{*}[-0.2ex]{\shortstack[l]{Accretion rate \\ ($10^{11}\,\mathrm{g\,s^{-1}}$)}}						
	& 0.25--2.0 &	\textbf{3.22} &	2.71 &	3.84 &	2.39 &	4.26 \\ 
	& 0.25--10.0 &	\textbf{11.55} &	8.63 &	15.09 &	7.04 &	17.40 \\ [1ex]
\hline						
						
   \multirow{2}{*}[-0.2ex]{\shortstack[l]{Accretion rate \\ ($10^{-14}\,\mathrm{M_{\odot}\,yr^{-1}}$)}}						
	& 0.25--2.0 &	\textbf{0.51} &	0.43 &	0.61 &	0.38 &	0.68 \\ 
	& 0.25--10.0 &	\textbf{1.83} &	1.37 &	2.39 &	1.12 &	2.76 \\ [1ex]

\hline
 \hline
    \end{tabular}
    \\
    \justifying
    \vspace{15pt}
    \noindent  
\end{table*}







\bsp	
\label{lastpage}
\end{document}